\documentclass[fleqn,usenatbib]{mnras}

\usepackage[usenames,dvipsnames]{xcolor}
\usepackage{hyperref}
\hypersetup{
    colorlinks = true,
    citecolor = {MidnightBlue},
    linkcolor = {BrickRed},
    urlcolor = {BrickRed}
}

\usepackage{newtxtext,newtxmath}
\usepackage[T1]{fontenc}
\usepackage{ae,aecompl}

\usepackage{graphicx}	
\usepackage{amsmath}	
\usepackage{balance}
\usepackage{multirow}

\usepackage{geometry}
\newcommand{\be}{\begin{equation}}
\newcommand{\ee}{\end{equation}}
\newcommand{\bea}{\begin{eqnarray}}
\newcommand{\eea}{\end{eqnarray}}

\title[Dark energy in the matter-dominated era]{Searching for dark energy in the matter-dominated era}

\author[P. Bull et al.]{
Philip Bull,$^{1,2}$\thanks{E-mail: p.bull@qmul.ac.uk}
Martin White,$^{3}$
and An\v{z}e Slosar$^{4}$
\\
$^{1}$Astronomy Unit, Queen Mary University of London, Mile End Road, London, E1 4NS, UK\\
$^{2}$Department of Physics \& Astronomy, University of the Western Cape, Cape Town 7535, South Africa\\
$^{3}$Department of Astronomy, University of California Berkeley, Berkeley, CA 94720, USA\\
$^{4}$Brookhaven National Laboratory, Physics Department, Upton, NY 11973, USA
}

\date{Accepted XXX. Received YYY; in original form ZZZ}

\pubyear{2019}

\begin{document}
\label{firstpage}
\maketitle

\begin{abstract}
Most efforts to detect signatures of dynamical dark energy are focused on late times, $z \lesssim 2$, where the dark energy component begins to dominate the cosmic energy density. Many theoretical models involving dynamical dark energy exhibit a ``freezing'' equation of state however, where $w \to -1$ at late times, with a transition to a ``tracking'' behaviour at earlier times (with $w \gg -1$ at sufficiently high redshift). In this paper, we study whether large-scale structure surveys in the post-reionisation matter-dominated regime, $2 \lesssim z \lesssim 6$, are sensitive to this behaviour, on the basis that the dark energy component should remain detectable (despite being strongly subdominant) in this redshift range given sufficiently precise observations. Using phenomenological models inspired by parameter space studies of Horndeski (generalised scalar-tensor) theories, we show how existing CMB and large-scale structure measurements constrain the dark energy equation of state in the matter-dominated era, and examine how forthcoming galaxy surveys and 21cm intensity mapping instruments can improve constraints in this regime. We also find that the combination of existing CMB and LSS constraints with DESI will already come close to offering the best possible constraints on $H_0$ using BAO/galaxy power spectrum measurements, and that either a spectroscopic follow-up of the LSST galaxy sample (e.g. along the lines of MegaMapper or SpecTel) or a Stage 2/PUMA-like intensity mapping survey, both at $z \gtrsim 2$, would offer better constraints on the class of dark energy models considered here than a comparable cosmic variance-limited galaxy survey at $z \lesssim 1.5$.
\end{abstract}



\section{Introduction}

Dark energy (DE) is generally considered to be a late-time phenomenon. According to current observational constraints \citep[e.g.][]{2018arXiv180706209P}, it only becomes an appreciable fraction of the cosmic energy density at redshifts of $z \lesssim 1$ or so, and begins to dominate over matter at $z \lesssim 0.3$. As such, most current observational studies -- primarily using supernova or galaxy surveys combined with CMB data -- focus on characterising dark energy at $z \lesssim 1$, while forthcoming large-scale structure surveys will extend their reach out to $z \approx 2$.

Higher redshifts are comfortably in the matter-dominated regime, where the kind of dark energy needed to cause late-time cosmic acceleration can only have a minor effect on the cosmic expansion rate, and where precision observational probes are scarce (typical galaxies and supernovae are too faint to easily detect in large numbers at such large distances). A notable exception is the Lyman-$\alpha$ forest, which has been used to measure the baryon acoustic oscillation (BAO) scale at $z \sim 2.4$ with a precision of a few percent \citep{2019arXiv190403430B}, but in practice the $z \gtrsim 2$ regime is beyond the effective reach of most current observational techniques.

While dark energy is sub-dominant at higher redshifts, this is not to say that it is phenomenologically uninteresting in this regime. So-called ``early dark energy'' models have long been pursued in the literature for example \citep[e.g.][]{2006APh....26...16L, 2006JCAP...06..026D, 2009JCAP...04..002X, 2011PhRvD..83l3504C, 2011PhRvD..83l3526M, 2013PhRvD..87h3009P, Poulin:2018cxd, Hill:2020osr, Ivanov:2020ril}. These posit a period of increased dark energy density at high redshifts, which is generally achieved by dynamically adjusting the equation of state, $w(z)$, to larger (less negative) values for some period. The mechanisms for achieving this adjustment vary, but for scalar field DE models typically include tuning the shape of the scalar field potential to allow temporary deviations from the slow-roll regime \citep[e.g.][]{1999PhRvD..59l3504S, 2018JCAP...08..009B}, or introducing couplings to matter or other fluids that modify the kinetic term of the scalar \citep[e.g.][]{DeFelice:2010pv, 2018arXiv180908735K}

While some models are specifically designed to give rise to early dark energy effects, recent work suggests that such phenomena might actually be reasonably generic, depending on the redshift range of interest. Horndeski models are the most general class of single scalar field theories in 4 dimensions, with at most second-order derivatives in the field \citep{Horndeski:1974wa, Deffayet:2011gz}. At linear order in cosmological perturbations, theories within this class can be fully specified by 4 arbitrary functions of time, subject to a set of physical viability conditions. In \citet{2017PhRvD..96h3509R}, theoretical priors on the dark energy equation of state were calculated by parametrising these functions in a broad way, Monte Carlo sampling the function coefficients, and then applying the physical viability conditions to reject unphysical models. Under the assumptions of this analysis, the resulting prior on $w(z)$ shows a tendency for the models to exhibit a broad `tracking'-type behaviour, where $w(z)$ tracks $w \simeq 0$ at high redshift, but transitions to a Cosmological Constant-like $w \simeq -1$ at low redshift. \citep[Note that some subclasses of Horndeski that exhibit tracking behaviours are disfavoured observationally however;][]{2013PhRvD..87j3511B, 2018arXiv180908735K}.

As long as the equation of state only differs significantly from $w=-1$ at higher redshifts, where there are few direct constraints on the expansion rate and distance-redshift relation, the effects on well-constrained quantities such as the distance to the CMB can generally be compensated by small shifts in other cosmological parameters, making these models difficult to distinguish from $\Lambda$CDM. Early dark energy does affect other observables however, for example by introducing a correction to the number of relativistic degrees of freedom, $N_{\rm eff}$, inferred from the CMB \citep{2011PhRvD..83l3504C, Hill:2020osr}, as well as affecting high-$z$ structure formation. These observables are at present relatively blunt instruments however, being degenerate with other effects (such as sterile neutrinos or differences in galaxy formation models respectively).

In this paper, we study the possibility of directly constraining the small modifications to the expansion history in the matter dominated regime that should arise as a result of a tracking DE equation of state. We begin by outlining a physical motivation for searching for DE phenomenology in the $2 \lesssim z \lesssim 6$ regime in Sect.~\ref{sec:demodels}, where we also discuss two parametrisations of the DE equation of state that can model tracking behaviours. We set out our hybrid observational parameter estimation and forecasting method in Sect.~\ref{sec:obs}, and present our results in Sect.~\ref{sec:results}. Finally, we conclude in Sect.~\ref{sec:conclusions}. In what follows, we adopt the best-fit parameters of the \citet{2018arXiv180706209P} {\tt base\_plikHM\_TTTEEE\_lowl\_lowE} $\Lambda$CDM analysis as our fiducial cosmology, with $h=0.6727$, $\Omega_{\rm M}=0.3166$, $\Omega_{\rm b}=0.04941$, $\sigma_8 = 0.8120$, and $n_s = 0.9649$.

\section{Dark energy models}
\label{sec:demodels}

In this section, we discuss the theoretical landscape of scalar field dark energy models, including recent work on defining generic priors on the equation of state. We then define two models for the equation of state: one based on a simple quintessence model (the Mocker model), and another based on a phenomenological parametrisation of the Horndeski class of models.

\subsection{Theoretical priors on the equation of state}

It has recently been rediscovered that a very general class of single scalar field models exists that can be parametrised (in the cosmological weak-field limit) by only a handful of arbitrary functions of time \citep{DeFelice:2010pv}. The Horndeski class encapsulates a large fraction of the scalar field dark energy and modified gravity models that have previously been studied in the literature, and organises them into subclasses according to which of these functions (which describe the time-dependent couplings of a small number of allowed operators in the action) take on non-trivial values. Different parametrisations of these models exist \citep[e.g.][]{Gubitosi:2012hu, Baker:2012zs, 2013JCAP...08..010B, 2014IJMPD..2343010G, 2014JCAP...07..050B} which allow background and linear perturbative expressions to be calculated with relative ease. There is little theoretical guidance on how the arbitrary functions of time should be chosen in these parametrisations however, beyond reproducing the behaviours of various specific scalar field models for which full (i.e. non-perturbative) actions have been written down, and applying a set physical viability conditions that prevent various instabilities from occurring.

Despite the arbitrary nature of the coupling functions, the fact that the scalar field must follow certain equations of motion, obey certain symmetries (defined by the allowed operators in the action), and respect physical viability conditions, imposes non-trivial structure in the behaviour of the theories. In other words, while the coupling functions are arbitrary, the possible dynamical behaviours of the theories are not. Several recent studies have taken these ingredients, along with very broad parametrisations of the arbitrary functions and broad observational priors, and performed Monte Carlo studies to establish theoretical priors on the dynamics of the Horndeski class and the resulting observable implications \citep{Perenon:2015sla, 2017PhRvD..96h3509R, Espejo:2018hxa, Gerardi:2019obr}; see also \cite{2012JCAP...02..048C}.

These studies find that particular functional forms of the equation of state are often preferred, mostly those exhibiting freezing-type behaviours (i.e. $w \to -1$) at low redshift, and a smooth transition to tracking-type behaviours at high redshift. While this does not necessarily rule-out more baroque forms of the equation of state (e.g. with oscillations, or sharp features), the implication is that significantly more tuning of the coupling functions is required to achieve these particular behaviours.

There are several reasons for the emergence of these apparently preferred functional forms. For minimally-coupled quintessence models, the equation of state is
\be
w = \frac{\frac{1}{2}\dot\phi^2 - V(\phi)}{\frac{1}{2}\dot\phi^2 + V(\phi)} = \frac{\epsilon - 1}{\epsilon + 1},
\ee
where we have defined $\epsilon$ to be the ratio of the kinetic term to the potential energy. To avoid a phantom quintessence ($w < -1$), we must have $\epsilon(a) \ge 0$ for all $a$. Observational constraints require $w \ll -{1/3}$ at late times, which implies $0 \le \epsilon \ll {1/2}$ around $a \simeq 1$ (i.e. the scalar field must be slowly rolling at late times). For slow, smooth, monotonic evolution of the equation of state satisfying both bounds one can start with $\epsilon > \epsilon(a=1)$ and have it decrease with time (a freezing model), or start with $0 \le \epsilon \lesssim \epsilon(a=1)$ and have it slowly increase or stay the same (a thawing model). For the models and assumed priors considered in \citet{2017PhRvD..96h3509R}, there are many more ways of achieving the former behaviour than the latter and so freezing models tend to be preferred prior to any constraints from data. The tendency towards a tracking behaviour at earlier times is due to a combination of this plus a similar bound; models that have $w > 0$ for more than a short period at early times are likely to either collapse or produce unrealistic abundances of matter and radiation, and so the equation of state at early times is essentially restricted to the range $-1 \lesssim w \lesssim 0$.

\begin{figure*}
	\includegraphics[width=\columnwidth]{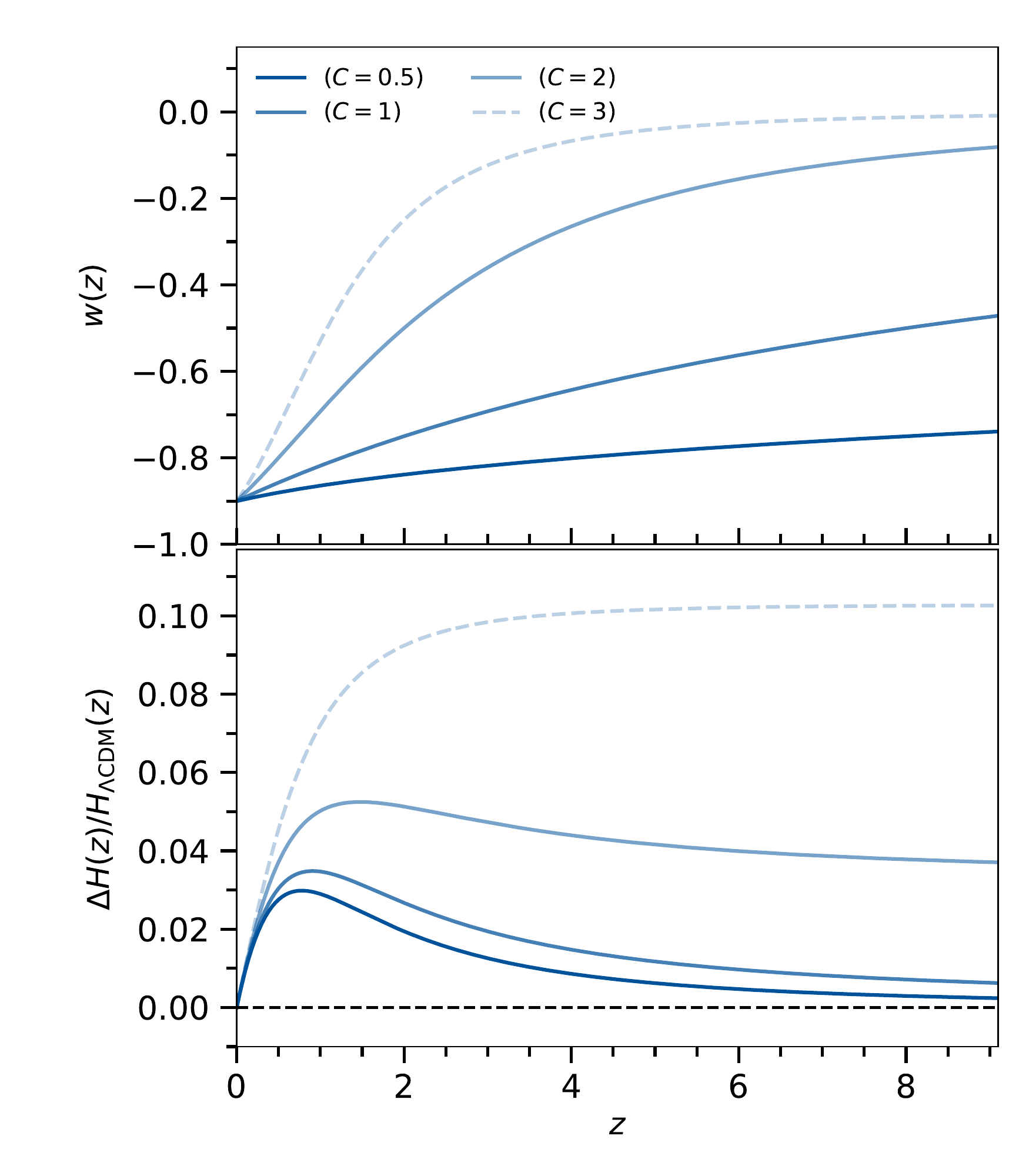}
    \includegraphics[width=\columnwidth]{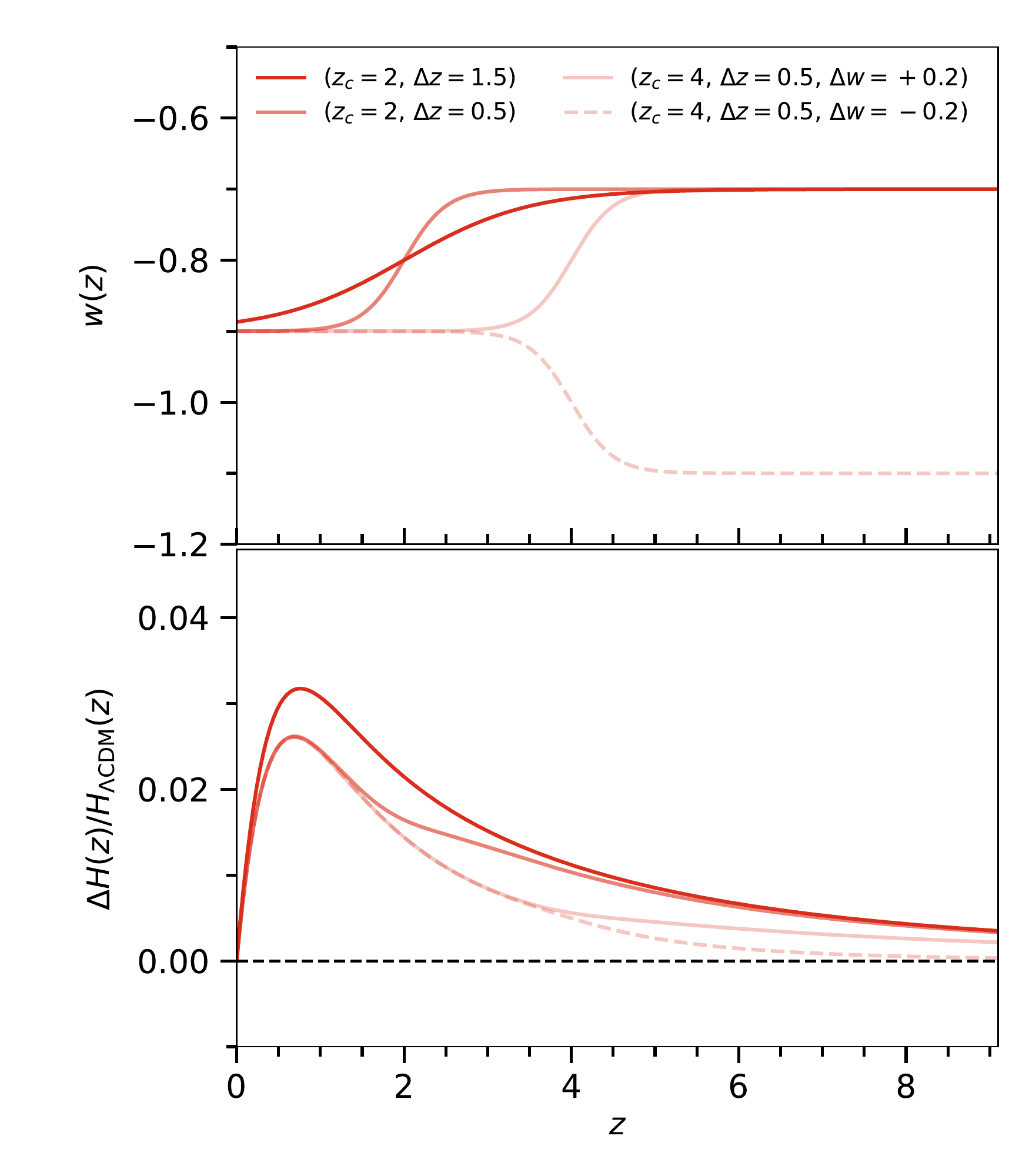}
    \caption{Equation of state of dark energy, $w(z)$, and fractional deviation of expansion rate from the fiducial $\Lambda$CDM value (where $\Delta H(z) = H(z) - H_{{\rm \Lambda}{\rm CDM}}(z)$) for a few examples of Mocker models (left panels) and phenomenological Tracker models (right). All models have $w_0 = -0.9$; the Tracker models also fix $\Delta w = w_\infty - w_0 = 0.2$ unless otherwise stated. For these relatively strong deviations from $\Lambda$CDM, there is a few-percent change in $H(z)$ that decays with a non-trivial redshift dependence.}
    \label{fig:model_wzhz}
\end{figure*}

Similar arguments apply to more general scalar field models, although this time the range of possible behaviours is broader due to the existence of new coupling terms. In such non-minimally coupled models, different fluids can interact and transfer energy between one another, and so the effective equation of state of the dark energy fluid can pass through $w=-1$ without any problem. In these models, the tracking behaviour arises to compensate for changes in the gravitational coupling strength (which is now an arbitrary function of time). Since the matter and radiation energy density are observationally well-constrained at early and late times, changes in the effective Newton's constant, $G_{\rm eff}$, that enhance or suppress their abundances must be compensated by the scalar field. The result is that only scalar field models that track the dominant component of the energy density can straightforwardly satisfy observational constraints \citep{2017PhRvD..96h3509R}.

While the discussion above tries to establish some `generic' behaviours of scalar field dark energy models that we can try to target, it is worth keeping in mind that such statements about the `size' of regions in model space are subject to a type of measure problem, in that we don't have a unique way to specify probability densities over the relevant model space. Physical viability conditions of the type applied by \cite{2017PhRvD..96h3509R} are useful because they can at least excise regions of the model space that are unphysical, although even these are not definitive; pathologies can sometimes be cured or pushed outside the domain of validity of the theory \citep[e.g. by breaking Lorentz invariance;][]{Konnig:2016idp}, making some `unphysical' theories viable again. Dynamical systems arguments, such as finding attractor solutions, are also far from watertight, as they are also affected by the ambiguity in the measure on the space of initial conditions (and whether those initial conditions are specified at early times and evolved forward in time, or vice versa).

As such, we are unable to make definitive claims about where it would be most `likely' to find interesting dark energy phenomenology given the set of all viable models; we can merely point at regions of model space that have interesting properties, subject to a particular set of assumptions. Hence, in this paper, we propose that tracking behaviours are sufficiently well-motivated from a theoretical perspective to consider targeting these models and their phenomenology observationally. We refrain from making any stronger statements about what a failure to observe such behaviours would imply for the viability of dynamical dark energy theories though; it will almost always be possible to come up with `designer' models that fit any particular observed expansion history. This picture is complicated somewhat once constraints on the growth of structure (particularly on non-linear scales) is also accounted for, but we will not consider them here.

\subsection{Mocker models}

As a specific example of a quintessence model with tracking behaviour, we consider the Mocker model discussed in \citet{2006APh....26...16L, 2006PhRvD..73f3010L}. This is a toy model, constructed to give a simple freezing-type behaviour without making any particular reference to physically-motivated scalar field models. It is defined by the relation $w^\prime = C w\,(1 + w)$, where $C$ is a constant, and $^\prime \equiv d/d\log a$. The solution for the equation of state is
\be
w(a) = \left [ \left ( \frac{1+w_0}{w_0} \right ) a^{-C} - 1 \right ]^{-1}
\ee
where $w(a=1) = w_0$ is a free parameter. The equation of state tends to a matter-like behaviour, $w\to 0$, at early times, while behaving as an accelerating fluid at late times, with the matter-dominated behaviour lasting for longer the larger the value of $C$. Setting $w_0=-1$ recovers the cosmological constant, regardless of the value of $C$.

The Mocker model is a minimally-coupled quintessence model, meaning that there are no other couplings to the matter sector beyond the gravitational interaction. The onset of tracking behaviour therefore depends only on the choice of $C$, i.e. a tuning that must be applied to the model, rather than through any physical coupling to other forms of energy. Choosing lower values of $C$ pushes the transition to a matter-like equation of state to earlier times. The value of $C$ is therefore bounded from above by observations, which support an accelerating fluid at low redshift. The lack of any non-minimal couplings constrains the equation of state to always remain on one side of the `phantom divide' ($w=-1$), since crossing it would give rise to perturbative instabilities. In what follows, we only consider models with $w_0 \ge -1$. Example Mocker model equations of state are shown in Fig.~\ref{fig:model_wzhz} (left panels).

\subsection{Phenomenological Tracker models}

As discussed above, generalised scalar field models admit non-minimal couplings that allow a broader range of behaviours at late times, whilst still preferring a tracking behaviour at earlier times. While a zoo of models can be constructed with all kinds of complex behaviour at late times, depending on the exact nature of the coupling and the effective scalar field potential, Monte Carlo studies of models with smooth, non-fine tuned couplings suggest that a relatively smooth transition in the equation of state from early to late times is quite typical \citep{2017PhRvD..96h3509R}. We propose a simple phenomenological model for $w(z)$ that exhibits this smooth transition behaviour while limiting the number of additional free parameters. Similar transitioning equation of state models have been considered widely in the literature however, with a variety of physical motivations and functional forms for the transition \citep[e.g.][]{1999PhRvD..59l3504S, 2000PhRvD..62h1302U, 2002MNRAS.336.1217B, 2006PhRvD..74h6009N, 2008PhRvD..78b3526L, 2018JCAP...08..009B}. We adopt the following 4-parameter model:
\be \label{eq:tracker}
w(z) = w_0 + \frac{1}{2} (w_\infty - w_0) \left ( 1 + \tanh \left ( \frac{z - z_c}{\Delta z} \right ) \right ).
\ee
This allows a Tracker-like behaviour at high redshift, where $w \to w_\infty \approx 0$, and the necessary accelerating behaviour at low redshift, where $w \to w_0 \approx -1$. There is a smooth interpolation between these two regimes, with a transition redshift set by $z_c$, and a transition width set by $\Delta z$.

This model also allows freezing and thawing behaviours, and has $w=-1$ as a special case. It can approximate the well-known CPL parametrisation, $w \approx w_0 + w_a(1 - a)$, when $|z - z_c| \ll \Delta z$. We also allow $w_0 < -1$, in recognition of the fact that the effective equation of state has no restriction on crossing the phantom divide.

\section{Observational constraints and forecasts} \label{sec:obs}

In this section, we examine current and future constraints on the representative tracking dark energy models discussed above. Our focus is on purely background-level constraints, i.e. those that use observables that primarily depend on the background expansion and geometry of the Universe, rather than the growth of perturbations. As such, we do not include redshift-space distortions or weak gravitational lensing observables, even though most of the experiments we consider are capable of measuring them, and they are both known to be strongly constraining of (e.g.) modified gravity scenarios. The main reason for this choice is the lack of a joint model of the prior on the equation of state and the growth rate. While quite general priors exist for these individually (see the discussion above about \cite{2017PhRvD..96h3509R} for $w(z)$, and \cite{Perenon:2015sla} for the growth rate), we are not aware of a joint prior that would properly enforce consistency between the two. Since constructing such a prior for general scalar field models is beyond the scope of this work, and building one from the handful of specific models discussed above would make our analysis much less general, we choose to focus only on background observables.

In the following subsections, we discuss the background observables that we do use, and how current and future surveys, and various combinations of them, are expected to impact constraints on the dark energy density at intermediate to high redshift in the post-reionisation regime.

\subsection{Existing constraints on distance measures}

At present, the most precise measurements of background quantities related to dark energy at late times come from spectroscopic galaxy surveys, which primarily target the baryon acoustic oscillation (BAO) feature in the galaxy power spectrum. The BAO feature can be decomposed into separate radial and transverse parts in order to measure the quantities
\bea
D_H(z) &\equiv& {c / H(z)} \\
D_M(z) &\equiv& (1+z)\, D_A(z)
\eea
respectively, or measured as a spherical average,
\be
D_V(z) \equiv \left (z\, D_H(z)\, D_M^2(z) \right )^\frac{1}{3}.
\ee
These quantities are often scaled by the sound horizon at the drag epoch, $r_d = r_s(z_{\rm drag})$, i.e. the physical BAO scale that we assume is known with significantly higher precision from the CMB. The radial and transverse distances $D_H$ and $D_M$ have also been measured by Lyman-$\alpha$ forest observations. A compilation of existing measurements from both types of survey is shown in Table~\ref{table:distances}.

\begin{table}
\begin{tabular}{llrrr}
\hline
{\bf Survey} & {\bf Type} & {\bf Redshift} & {\bf Distance} \\
\hline
6dFGS & $D_V/r_d$ & 0.11 & $3.047 \pm 0.137$ Mpc \\
MGS & $D_V/r_d$ & 0.15 & $4.480 \pm 0.168$ Mpc \\
BOSS LOWZ & $D_V/r_d$ & 0.32 & $8.467 \pm 0.167$ Mpc \\
\hline
\multirow{3}{*}{BOSS CMASS} & $D_M/r_d$ & \multirow{3}{*}{0.57} & $14.945 \pm 0.210$ Mpc \\
 & $D_H/r_d$ &  & $20.75~ \pm 0.73~\,$ Mpc \\
 & Corr. coeff. &  & $-0.52~~~~~~~~~~$ \\
\hline
\multirow{3}{2.05cm}{eBOSS Ly$\alpha$ auto+QSO} & $D_M/r_d$ & \multirow{3}{*}{2.34} & $37.0~ \pm 1.3~\,$ Mpc \\
 & $D_H/r_d$ &  & $9.00 \pm 0.22$ Mpc \\
 & Corr. coeff. &  & $-0.40~~~~~~~~~~$ \\
\hline
\end{tabular}
\caption{Distance measurements used in this paper, taken from \citet{2015PhRvD..92l3516A} (first 5) and \citet{2019arXiv190403430B} (last 2). The errorbar on the eBOSS Ly$\alpha$ $D_M$ measurement has been symmetrised.}
\label{table:distances}
\end{table}

We take measurements from each survey and redshift bin to be independent, and construct a Gaussian likelihood of the form
\be
\log\mathcal{L}_i(\vec{\theta}) \sim -\frac{1}{2} [\vec{D}(z_i, \vec{\theta}) - \vec{D}_{i}]^T C^{-1} [\vec{D}(z_i, \vec{\theta}) - \vec{D}_{i}] \label{eq:logL}
\ee
for a redshift bin centred on $z_i$, where $\vec{D} = (D_M, D_H)$ or $(D_V)$ depending on the survey, and $C$ is the appropriate covariance matrix for $\vec{D}_i$. Theoretical predictions for $\vec{D}(z_i)$ can be obtained by adopting one of the parametrised functional forms for $w(z)$ given above, solving for the energy density of the dark energy fluid,
\be
\rho_{\rm DE}(z) = \rho_{{\rm DE}, 0}\exp \left ( -3\int [1 + w(a)] \, d\log a \right ),
\ee
and then inserting this into $H(z)$ to calculate the relevant distance measure.

\begin{figure*}
	\includegraphics[width=1.0\columnwidth]{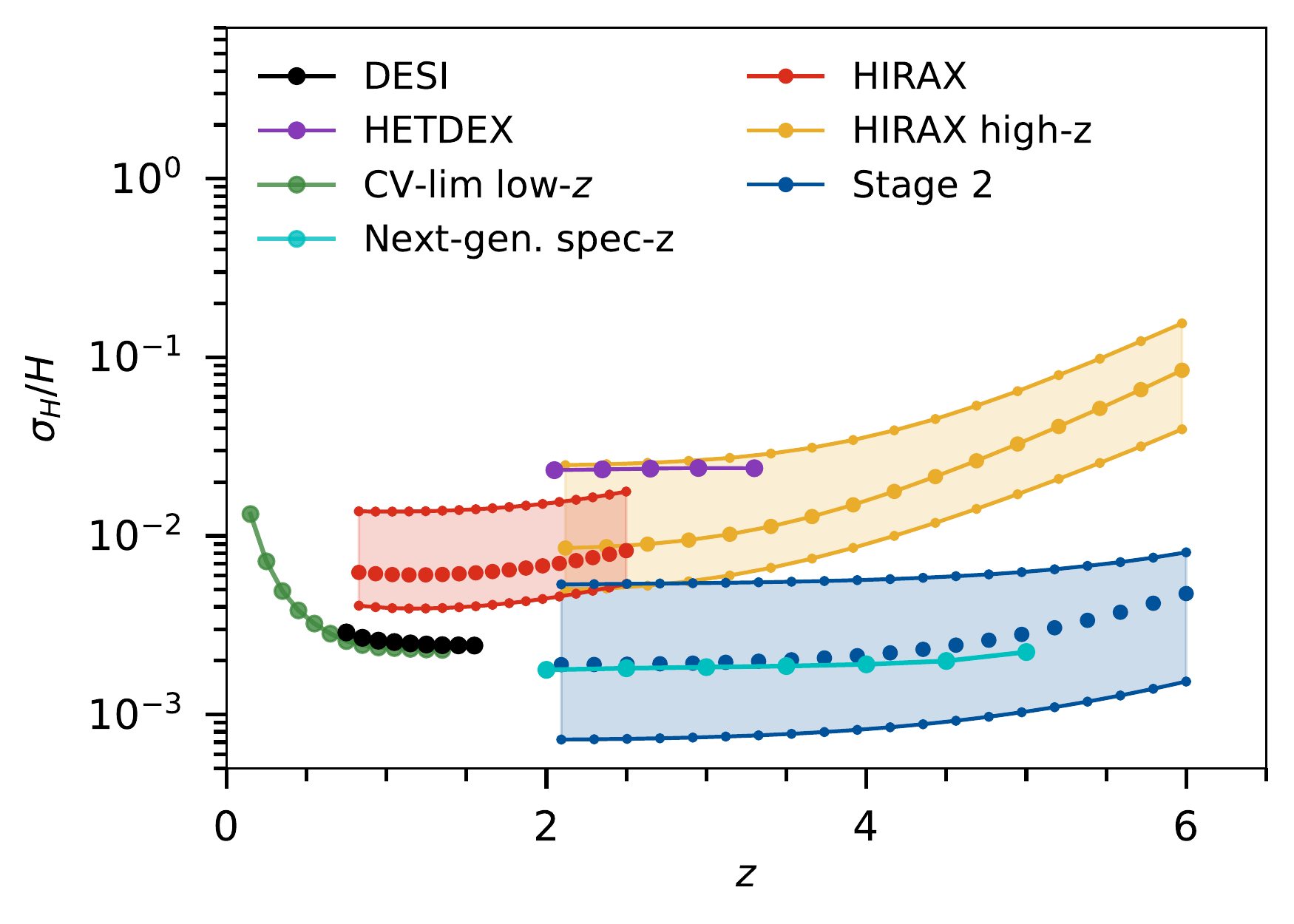}
    \includegraphics[width=1.0\columnwidth]{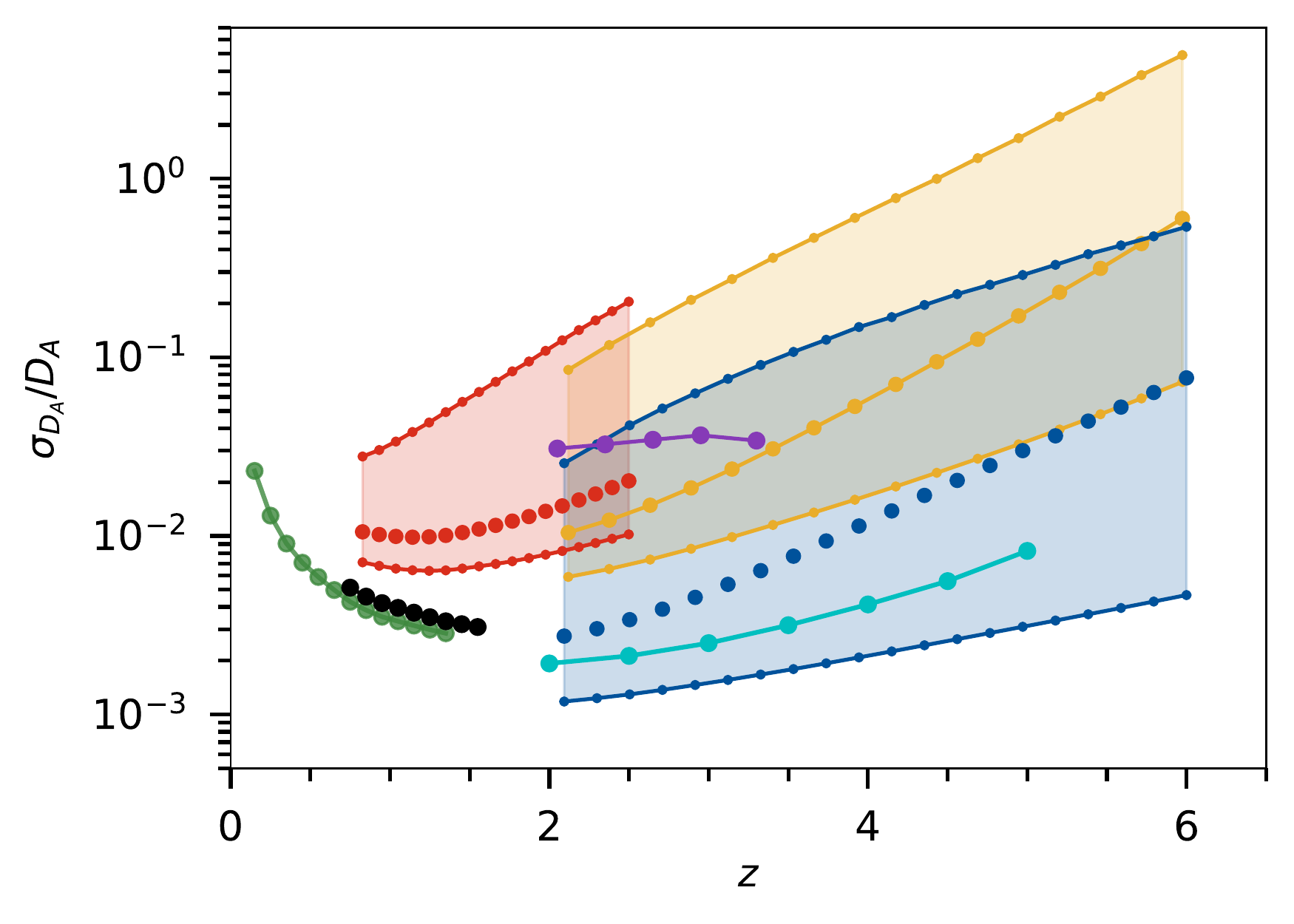}
    \caption{Forecast fractional errors (68\% CL) on $H(z)$ (left) and $D_A(z)$ (right) for DESI \citep{Aghamousa:2016zmz}, HETDEX \citep{Hill:2008mv}, HIRAX \citep{Newburgh:2016mwi}, a high-redshift version of HIRAX \citep[c.f.][]{2018JCAP...05..004O}, a Stage 2 intensity mapping experiment \citep{Ansari:2018ury}, a hypothetical cosmic variance-limited low-redshift galaxy survey, and a next-generation spectroscopic galaxy survey \citep[c.f.][]{Ferraro:2019uce}. For the IM experiments, results for three foreground removal types are shown: (upper points) horizon wedge removal; (middle points) $3\times$ primary beam wedge removal; (lower points) no foreground wedge. The IM experiments show the characteristic degradation in $\sigma_{D_{\rm A}}$ as their angular resolution decreases (and thermal noise increases) with redshift.}
    \label{fig:obs_errs}
\end{figure*}

\begin{table*}
\begin{tabular}{lrrrrr}
\hline
{\bf Experiment} & {\bf No. Dishes} & {\bf Redshift} & {\bf $S_{\rm area}$ $[{\rm deg}^2]$} & {\bf Dish Size} & {\bf $T_{\rm inst}$ [K]} \\
\hline
DESI            & --- & $0.7 - 1.7$ & 14,000 & --- & --- \\
HETDEX          & --- & $1.9 - 3.5$ & 420 & --- & --- \\
CV-lim. low-z   & --- & $0.1 - 1.5$ & 20,500 & --- & --- \\
Next-gen. spec-z & --- & $1.8 - 5.3$ & 14,000 & --- & --- \\
\hline
HIRAX           & 1024 & $0.8 - 2.6$ & 20,500 & 6m & 50 \\
HIRAX (high-z)  & 1024 & $2.0 - 6.1$ & 20,500 & 6m & 50 \\
Stage 2         & 65536 & $2.0 - 6.1$ & 20,500 & 6m & 50 \\
\hline
\end{tabular}
\caption{Assumed survey specifications. We have assumed DESI to be sample variance-limited in the given redshift range (and to have no sensitivity outside that range), and that each intensity mapping experiment spends a total of 10,000 hours on sky.}
\label{table:experiments}
\end{table*}

The BAO feature has been found to be particularly robust to systematic effects, hence its widespread use. Its use as a distance measure depends on having accurate knowledge of its physical scale however, which is set by the radius of the sound horizon during the drag epoch, $r_d$. This is constrained by \citet{2018arXiv180706209P} -- for the {\tt TT,TE,EE+lowE} data combination and a $\Lambda$CDM + $\Omega_{\rm K}$ model -- to be 
$r_d = 147.05 \pm 0.30$ Mpc
when the sum of neutrino masses is fixed to $\sum m_\nu = 0.06$ eV. The error on this quantity is only $0.2\%$, which is much smaller than the errors on typical distance measurements, and so for simplicity we fix it to the Planck best-fit value. We also fix the energy density of radiation (${\rm T}_{\rm CMB} = 2.725$ K), the sum of neutrino masses, $\sum m_\nu = 0.06$ eV, and the effective number of relativistic degrees of freedom $N_{\rm eff} = 3.046$.

In addition to low redshift constraints, we also include Planck CMB distance measurements at high redshift to act as an anchor point. For these, we construct a simplified likelihood involving only the relevant background quantities: $\Omega_b h^2$, $\Omega_c h^2$, and the distance to the CMB, $D_A(z_*) / r_s(z_*)$, in a similar way to `shift parameter' approaches that have been used in previous studies \citep[e.g.][]{2006ApJ...650....1W, Vonlanthen:2010cd, 2015PhRvD..92l3516A}. We do this by taking the Planck full-polarisation $\Lambda$CDM + $\Omega_{\rm K}$ MCMC chains \citep[{\tt base\_omegak\_plikHM\_TTTEEE\_lowl\_lowE};][]{2018arXiv180706209P}, deriving the three background quantities described above for each sample, and then marginalising over other parameters necessary to describe the CMB power spectrum (the normalisation and spectral index of the primordial scalar power spectrum, and optical depth to last scattering, plus the standard set of nuisance parameters). We then calculate the mean and covariance matrix for the three parameters of interest, and insert these into a Gaussian likelihood. We checked that the MCMC posteriors are well-approximated by a multi-variate Gaussian for these parameters. The rationale for using this particular combination of Planck data and cosmological model is that allowing $\Omega_{\rm K}$ to be a free parameter introduces a geometric degeneracy in the distance to the CMB in a similar way to dark energy models, which has the effect of relaxing the constraints on the other background parameters. We do not include CMB lensing information, which is also sensitive to changes in the growth history due to dark energy, which we do not model here.

\subsection{Fisher matrix predictions for future experiments} \label{sec:fisher}

For future experiments, we consider both spectroscopic galaxy surveys and 21cm intensity mapping surveys. We also relax the requirement to only consider information in the BAO feature, as if sufficient control over systematic effects can be achieved, future surveys will be able to use the broadband shape of the power spectrum to measure distances, therefore making use of many more Fourier modes and significantly improving their accuracy.

To obtain predictions for each survey, we use the method in \cite{Bull:2014rha} \citep[following][]{Seo:2007ns} and calculate the Fisher matrix,
\be
F_{ij}(z_n) = \frac{1}{2} V_{\rm bin} \int d^3k\, \frac{\partial \log P_{\rm tot}(\vec{k})}{\partial \theta_i} \frac{\partial \log P_{\rm tot}(\vec{k})}{\partial \theta_j},
\ee
where $V_{\rm bin}$ is the comoving survey volume of a redshift bin centred on $z_n$, $P_{\rm tot} = P_{\rm S} + P_{\rm N}$ is the total (signal plus noise) 3D power spectrum, and $\vec{\theta}$ is a set of cosmological and nuisance parameters to be marginalised over. The Fisher matrix for both spectroscopic galaxy surveys and intensity mapping surveys can be written in this form, but with the noise term taking on a non-trivial $\vec{k}$-dependence in the IM case to account for the limited angular resolution and filtering of foreground modes in these experiments. In both cases, the signal power spectrum is written as
\be
P(k, \mu, z) \propto (b(z) + f(z) \mu^2)^2\, P(k, z)\, e^{-k^2\mu^2 \sigma_{\rm NL}^2},
\ee
where the constant of proportionality is $1$ for galaxy surveys, and the brightness temperature squared, $T^2_b(z)$, for IM surveys. The first term in parentheses is a redshift-space distortion term, and the exponential term accounts for damping of the redshift-space power spectrum by incoherent peculiar velocities on small scales. We forecast for 4 parameters per redshift bin -- $D_M(z)$, $D_H(z)$, $f(z)$, $b(z)$ -- and marginalise over $\sigma_{\rm NL}$ with a fixed redshift dependence. The first two parameters are measured from the broadband shape of the redshift-space power spectrum (i.e. not just the BAO feature), while the latter two are marginalised to account for uncertainties in the bias and growth models. All other parameters are held fixed, and redshift bins are chosen to be sufficiently broad that they can be treated independently. Full details of the forecasting method are given in \cite{Bull:2014rha}.

With Fisher matrices for each redshift bin in hand, we construct an effective covariance for $D_M$ and $D_H$, marginalised over all of the other forecast parameters, and construct likelihoods of the same form as Eq.~\ref{eq:logL}. Instead of measured distances, we substitute $\vec{D}_i$ that take on their respective fiducial $\Lambda$CDM values. We do not add a noise realisation to this simulated data vector, so the best-fit model should always be the fiducial model for these experiments.

The specifications for each survey that we consider are shown in Table~\ref{table:experiments}. We have taken a simplified version of DESI, which we assume to be sample variance limited in the redshift range $z=0.7 - 1.7$. This is close to reality, except the relevant galaxy samples will extend slightly outside this range (albeit with lower galaxy number densities), and a low-redshift galaxy sample has been omitted. For the sake of comparison, we have included a hypothetical low-redshift spectroscopic galaxy survey that is sample variance-limited from $z=0.1-1.5$ over half the sky, with an assumed mean bias of $b(z) = \sqrt{1 + z}$. We have also included a next-generation high-redshift spectroscopic galaxy survey, following the `idealised' specification from \cite{Ferraro:2019uce}. This is an optimistic representation of what dedicated spectroscopic follow-up surveys of the final LSST 10-year galaxy sample may be able to achieve, e.g. see proposals such as FOBOS \citep{2019BAAS...51g.198B}, Maunakea Spectroscopic Explorer \citep{2019BAAS...51g.126M}, MegaMapper \citep{Schlegel:2019eqc}, and SpecTel \citep{2019BAAS...51g..45E}.

Turning to the intensity mapping experiments, the HIRAX specifications were adapted from \cite{Newburgh:2016mwi}, and we have also added a notional high-redshift version of the experiment, which would use the same dishes, correlator etc., but replace all of the receivers with lower-frequency versions operating at $200 - 475$ MHz with the same instrumental noise temperature. This is a sub-optimal design for this redshift range; in a more practical design, the baseline lengths and dish sizes should also be scaled up to counteract the decreasing angular resolution at lower frequency. For all of the intensity mapping experiments, we have assumed an effective survey area of half the sky, which is reasonable for drift scan telescopes if there is no need to mask out a substantial fraction of the Milky Way. We have also assumed parabolic dishes of diameter 6m for all experiments, and a constant instrumental noise temperature of $T_{\rm inst} = 50$K (which must be added to the sky temperature to give the total system temperature, $T_{\rm sys} = T_{\rm sky} + T_{\rm inst}$). 

\begin{table}
\centering
\begin{tabular}{|l|rr|}
\hline
{\bf Parameter} & {\bf Min.} & {\bf Max.} \\
\hline
$h$ & $0.5$ & $0.8$ \\
$\Omega_{\rm M}$ & $0.2$ & $0.4$ \\
$\Omega_{\rm b}$ & $0.01$ & $0.1$ \\
\hline
{\bf Mocker} & & \\
\hline
$w_0$ & $-1.0$ & $-0.1$ \\ 
$C$ & $1.3$ & $1.7$ \\
\hline
{\bf Tracking} & & \\
\hline
$w_0$ & $-2.0$ & $-0.1$ \\
$\Delta w$ & $-2.0$ & $2.0$ \\
$z_c$ & $-0.2$ & $10.0$ \\
$\Delta z$ & $0.01$ & $10.0$ \\
\hline
\end{tabular}
\caption{Priors on cosmological and dark energy model parameters used in the MCMC fits, all of which are assumed to follow a uniform distribution. We have defined $\Delta w \equiv w_\infty - w_0$.}
\label{table:priors}
\end{table}

For each of the intensity mapping experiments, we consider three different foreground mitigation scenarios. The most conservative (`horizon wedge') is when all modes that could possibly be affected by foregrounds\footnote{Foreground power can be scattered outside the wedge region by instrumental effects such as cable reflections, but we assume that any such effects have already been mitigated by the instrumental design.} are completely removed, resulting in a significant loss of signal Fourier modes too \citep{Thyagarajan:2015ewa}. The most optimistic (`no wedge') is when perfect foreground removal is assumed, allowing all signal modes to be recovered. Finally, an intermediate case (`$3\times$ PB wedge') is when only a smaller wedge region, corresponding to three times the angular size of the main lobe of the primary beam (i.e. including the main lobe and first couple of sidelobes), is removed.

Fig.~\ref{fig:obs_errs} shows the forecast fractional errors on $H(z)$ and $D_A(z)$ for all of the experiments we considered. The right-hand panel shows the characteristic degradation of constraints on $D_A(z)$ with redshift for the intensity mapping experiments, whose angular resolution decreases at lower frequencies.

\subsection{MCMC parameter estimation}
\label{sec:mcmc}

With the likelihoods for each of the current and future surveys in hand, we combine them in various sets (assuming independence), and use the {\tt emcee} affine-invariant ensemble sampler \citep{2013PASP..125..306F} to fit either the Mocker or Tracker models to the data. Our model fits involve 3 background parameters ($h, \Omega_{\rm M}, \Omega_{\rm b}$), plus 2 additional parameters for the Mocker models ($w_0, C$), and 4 for the Tracker models ($w_0, \Delta w, z_c, \Delta z$). The parameters and their priors are summarised in Table~\ref{table:priors}.

Some notes on the priors are necessary. First, for the tracking model, we chose a maximum transition redshift of $z_c = 10$ and a maximum transition width of $\Delta z = 10$. This restricts our analysis to an `interesting' region of parameter space where a transition (or hints of a transition) in the equation of state could be directly observed with large-scale structure experiments. Broader, and higher-redshift, transitions are observationally viable, but have little impact on the observables we are considering in this paper, and are essentially degenerate with one another -- particularly in terms of their low-redshift behaviour. It can be seen that Eq.~\ref{eq:tracker} has the limit $w(z) \to w_0$ as $z_c \to \infty$, making these models essentially indistinguishable from $w={\rm const.}$ models for our purposes.

For each model and combination of experiments, we ran chains of 80 walkers with 100,000 samples per walker, resulting in chains of 8 million samples. We thinned the chains stored on disk to keep only every 50th sample per walker. After performing convergence tests, we (conservatively) discarded the first 25\% of the remaining samples as burn-in, leaving 120,000 samples in total. Because our fiducial $\Lambda$CDM model does not have a transition feature in the equation of state (corresponding to $\Delta w = 0$), the region of the Tracker model parameter space being explored is one in which $z_c$ and $\Delta z$ are essentially unconstrained, and so the long length of the chains gives the walkers sufficient time to properly mix throughout the unconstrained subspace.

\section{Results}
\label{sec:results}

In this section we present the results of our MCMC parameter estimation runs on combinations of real and forecasted data. In all cases we include a likelihood based on the existing real data, denoted by `CMB + LSS'. For future experiments we also include forecasted data that assume a particular fiducial cosmology (and a `data' vector without any noise added, so that it perfectly aligns with the fiducial theoretical model).

\begin{figure}
	\includegraphics[width=1.0\columnwidth]{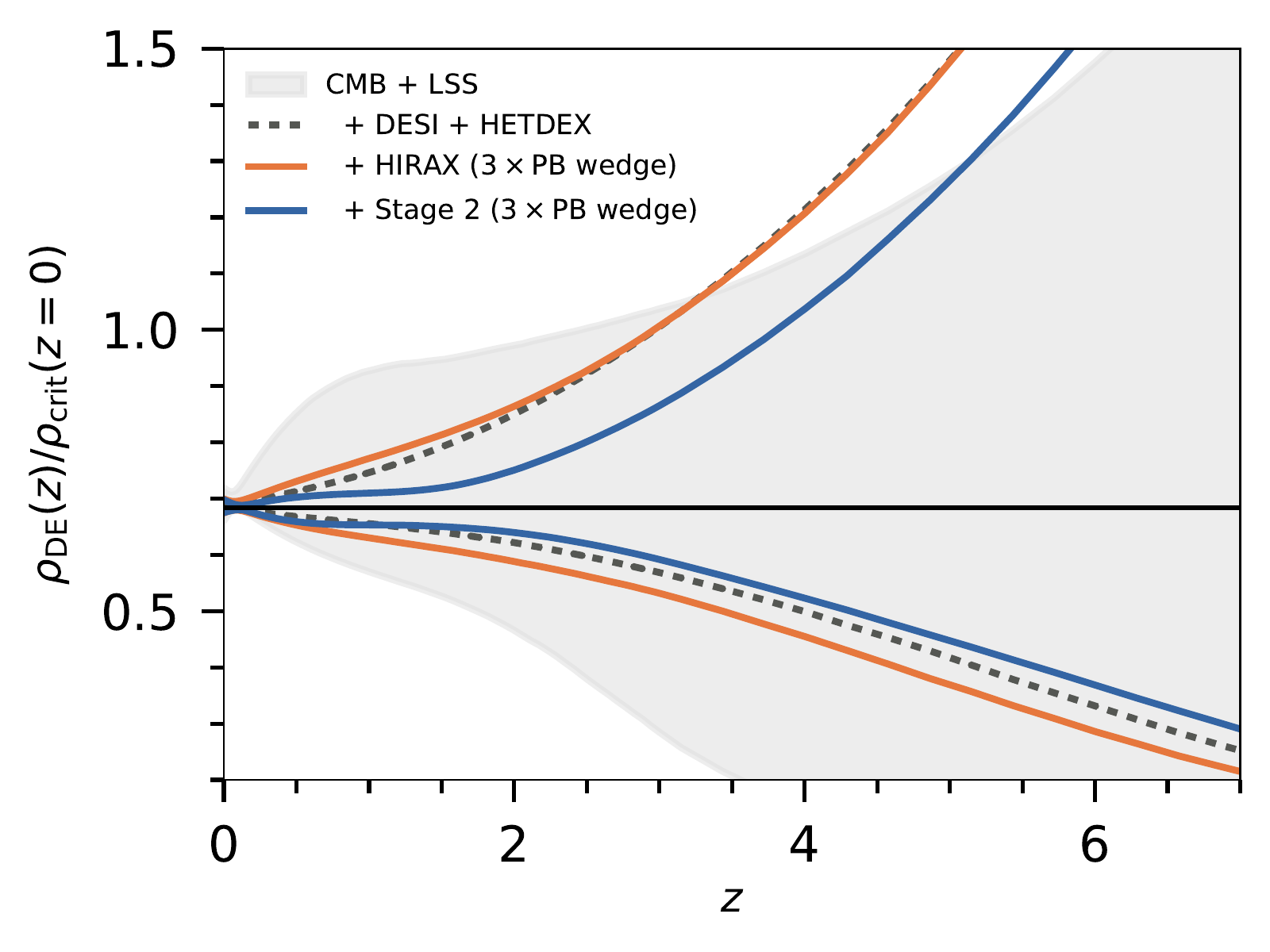}
    \caption{Forecast $95\%$ CL constraints on $\rho_{\rm DE}(z)$ in the Tracker model. The grey region shows the results for existing observations, a combination of CMB + large-scale structure constraints summarised in Table~\ref{table:distances}, while the solid/dashed lines show the combination of these existing data with forecasts for a selection of future experiments. The black horizontal line shows the fiducial $\Lambda$CDM model. Note that each line is for only CMB + LSS plus the experiment(s) listed in the legend, i.e. ``+ HIRAX'' should be read as ``CMB + LSS + HIRAX'', and does not include any other experiment.}
    \label{fig:omegade_galsurv}
\end{figure}

\begin{figure*}
	\includegraphics[width=1.0\columnwidth]{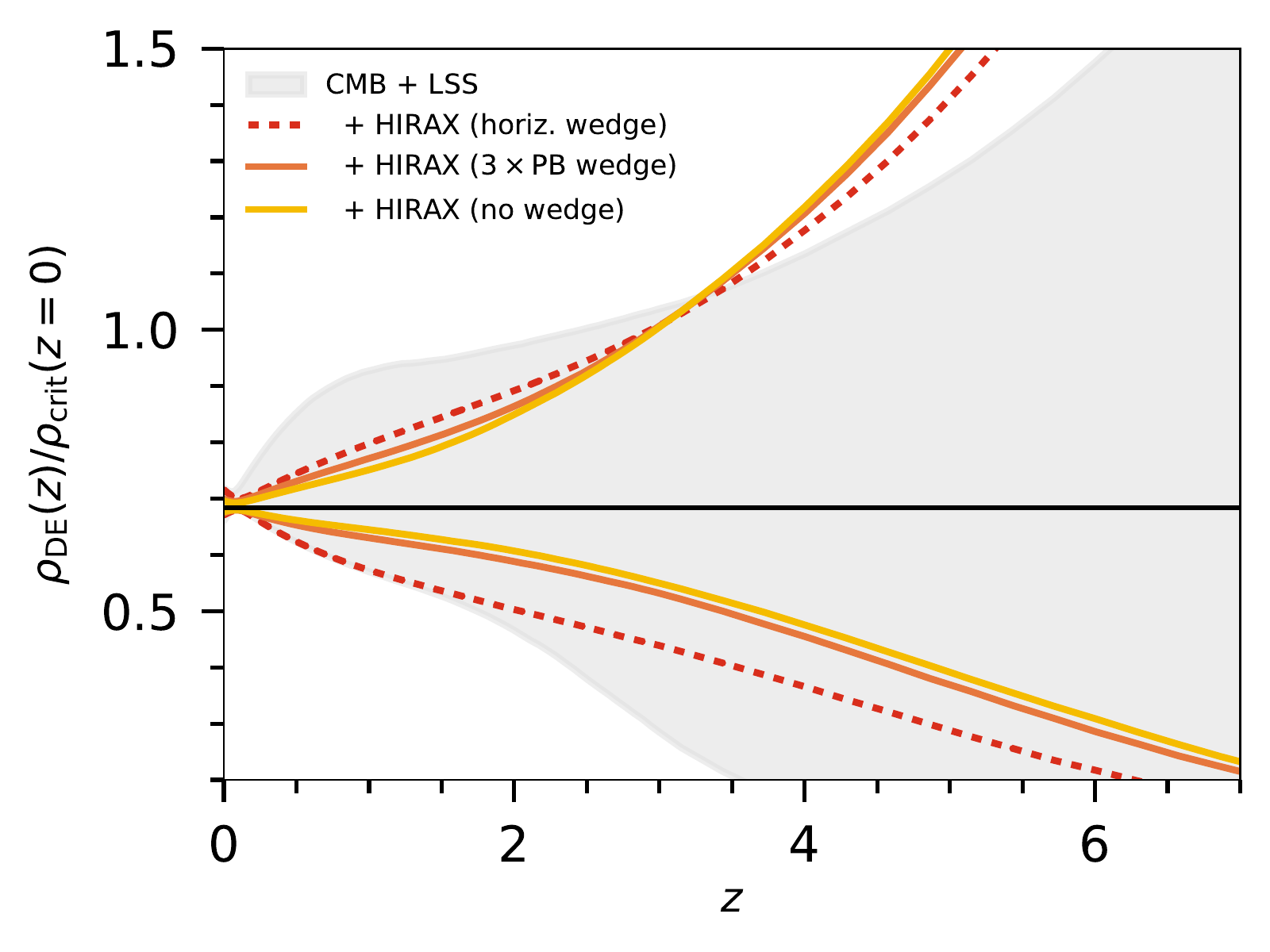}
	\includegraphics[width=1.0\columnwidth]{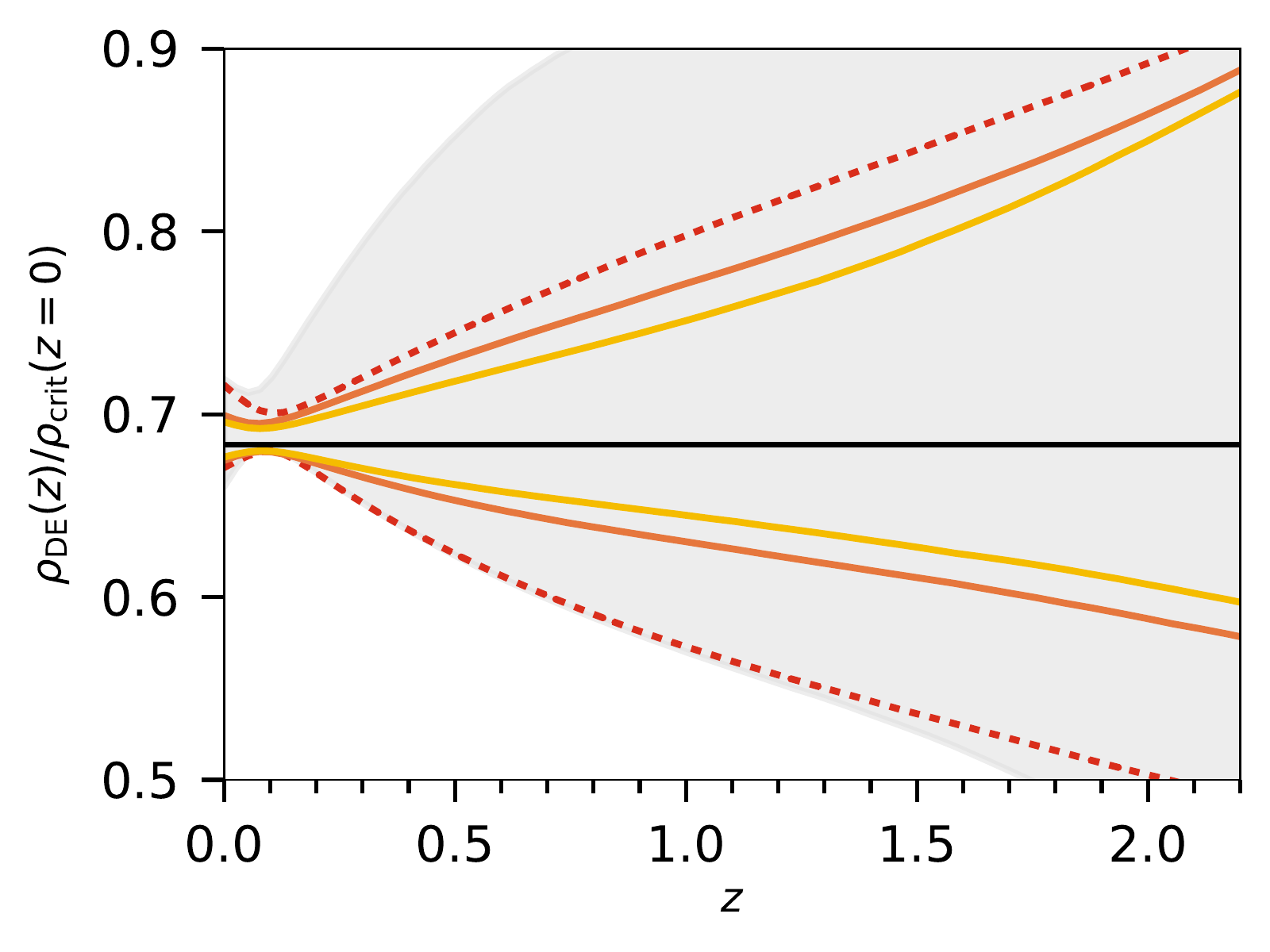}
    \caption{Forecast $95\%$ CL constraints on $\rho_{\rm DE}(z)$ in the Tracker model, from the combination of CMB + LSS with a HIRAX 21cm intensity mapping survey with different assumptions about foreground removal. ({\it Left panel:}) Constraints in the redshift range $z = 0 - 7$. ({\it Right panel:}) Detail at lower redshift.}
    \label{fig:omegade_hirax}
\end{figure*}

\subsection{Comparison of existing and future constraints}

Fig.~\ref{fig:omegade_galsurv} shows constraints on $\rho_{\rm DE}(z)$ for the Tracker model, normalised by the critical density at $z=0$. The grey region shows the results for existing `CMB + LSS' constraints: the combination of Planck CMB data, plus galaxy redshift survey and Lyman-$\alpha$ forest observations listed in Table~\ref{table:distances}. Also shown are the combination of these existing data with combined forecasts for two near-future galaxy redshift surveys (DESI and HETDEX); the HIRAX intensity mapping experiment (assuming moderately conservative $3\times$ primary beam wedge foreground removal); and a future Stage 2 intensity mapping experiment (also with $3\times$ PB wedge removal).

Within the context of the Tracker model, all of these experiments have broadly similar sensitivities to both growing and decaying dark energy densities into the past. The existing CMB + LSS constraints are asymmetric about the $\Lambda$CDM line however, translating to a slight preference for models where $\rho_{\rm DE}(z)$ grows into the past, but where a Cosmological Constant-like behaviour ($\rho_{\rm DE} = {\rm const.}$) is still well within the observational uncertainties. The other combinations of surveys also show a mild asymmetry, but without the `bump' feature in the contours seen for the CMB + LSS data around $z=1$.

A notable feature in Fig.~\ref{fig:omegade_galsurv} is the region beyond $z \approx 3.5$ where the bounds on $\rho_{\rm DE}(z)$ from the forecast constraints are broader than for the existing data. When adding forecasts for future experiments, we have assumed a particular fiducial $\Lambda$CDM model that is absolutely consistent between experiments, and so expect to recover that. The existing constraints, on the other hand, include all of the complexities of real data, including possible systematics and inconsistencies between datasets, which will result in deviations from the Planck best-fit fiducial model that we have assumed throughout the rest of our analysis. 
The reason for the non-overlapping region is mostly due to this difference between the fiducial model and the preferred model for the CMB + LSS dataset. Since $\rho_{\rm DE}(z)$ is a derived quantity, regions that appear to be `excluded' by CMB + LSS alone can appear to become viable again for CMB + LSS + other experiments. This is ultimately just an effect caused by the reweighting of different regions of the parameter space as new (forecast) constraints are added to the likleihood; for the sampled (non-derived) parameters themselves, there is no recovery of regions of the parameter space excluded by CMB + LSS, although there are significant shifts in where the bulk of the posterior lies (see Sect.~\ref{sec:cosmoparams}).

Given that we are assuming a particular fiducial model, the main point of comparison between different experiments in our study is therefore the size of the confidence regions rather than the recovered best-fit parameters. It can be seen from Fig.~\ref{fig:omegade_galsurv} that the combination of DESI and HETDEX with existing CMB + LSS constraints is already quite powerful, shrinking the confidence intervals across redshift by a factor of several compared to CMB + LSS alone (particularly for models where $\rho_{\rm DE}(z)$ is decaying into the past). The HIRAX results are not quite as constraining; this experiment produces a factor of $\sim\! 2$ worse constraints on $H$ and $D_A$ than DESI in the redshift range in which they overlap (see Fig.~\ref{fig:obs_errs}); outperforms HETDEX by a similar factor where they overlap; but results in only slightly broader confidence intervals on $\rho_{\rm DE}(z)$ than the combination of DESI + HETDEX. The difference between the DESI + HETDEX and HIRAX scenarios appears to be driven largely by the better low-redshift constraints from DESI, as adding DESI without HETDEX produces similar results.

Adding the Stage 2 intensity mapping experiment results in the best constraints out of the four scenarios shown in Fig.~\ref{fig:omegade_galsurv}. In the $3\times$ PB wedge foreground removal scenario, this experiment produces similar percentage-level constraints on $H$ and $D_A$ as DESI, but only at redshifts $z \gtrsim 2$. This produces a factor of $\sim 2$ improvement compared with the DESI + HETDEX constraints on $\rho_{\rm DE}(z)$ for models where the dark energy density grows into the past (compare the upper blue solid and black dashed lines in Fig.~\ref{fig:omegade_galsurv}), but a roughly similar constraint for models where it decays into the past (the lower blue and black dashed line in that figure). Referring back to Fig.~\ref{fig:model_wzhz} suggests a qualitative explanation for this behaviour. Models with a decaying dark energy density into the past have expansion rates that are practically indistinguishable from $\Lambda$CDM at higher redshifts, and so the high-$z$ bins of the Stage 2 survey add little additional information about them, while the lower-$z$ bins appear to offer a similar constraining power to DESI on the relevant parameters. Models where $\rho_{\rm DE}(z)$ is growing into the past, however, produce small but measurable differences in $H(z)$ that can be detected at higher $z$, resulting in a more significant improvement in the constraints from Stage 2 in this case.

Our conclusions from this section are that DESI (and HETDEX) will already be able to significantly improve constraints on Tracker-type dark energy models in the relatively short term, by surveying the lower-redshift regime that is generally considered the most obvious target for dark energy science. Experiments such as HIRAX that bridge the low- and higher-redshift regimes will not be quite as constraining, but are still competitive. In the medium term, however, higher-redshift experiments like Stage 2 look to be a better prospect for achieving significant improvements over DESI + HETDEX. (We will discuss in Sect.~\ref{sec:lowzhighz} how a futuristic cosmic variance-limited low-$z$ experiment compares with Stage 2.)

\begin{figure*}
	\includegraphics[width=1.0\columnwidth]{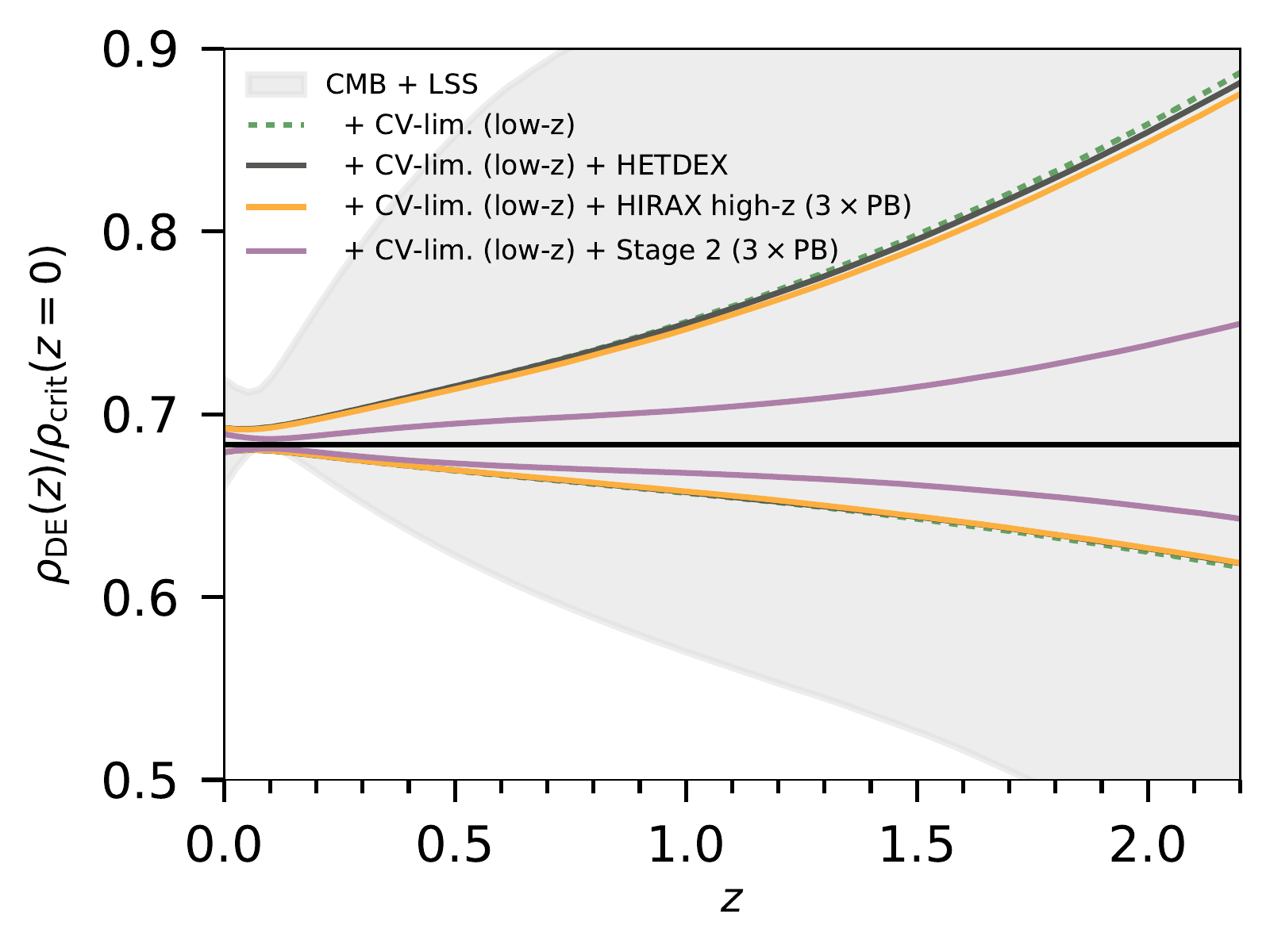}
	\includegraphics[width=1.0\columnwidth]{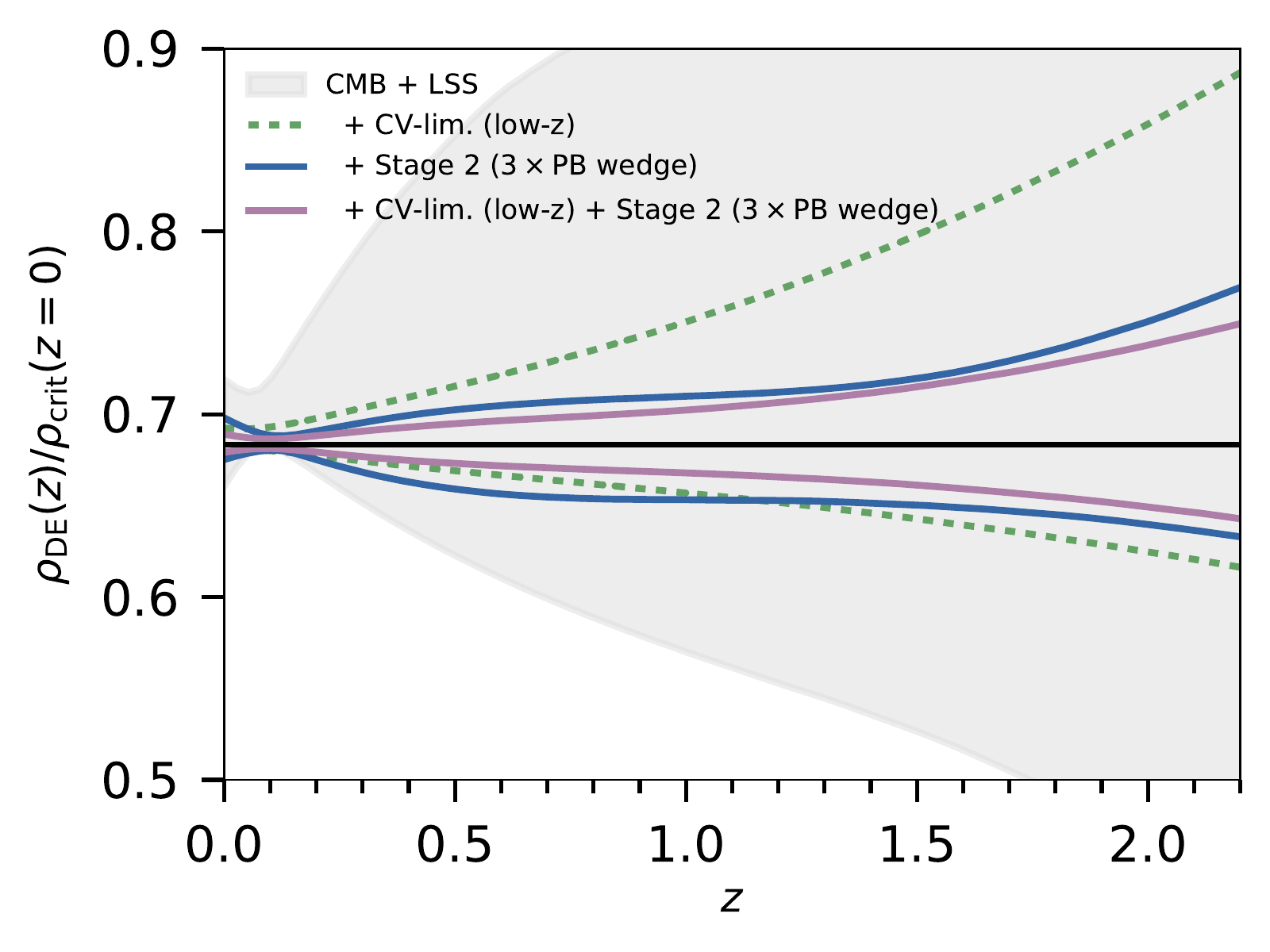}
    \caption{Forecast $95\%$ CL constraints on $\rho_{\rm DE}(z)$ in the Tracker model. ({\it Left panel:}) Constraints from a low-redshift cosmic variance-limited galaxy survey in combination with a variety of other experiments. The lines for the first three combinations of experiments all fall on top of one another, so are hard to distinguish in this plot. ({\it Right panel:}) Comparison of the CV-limited survey with the high-redshift Stage 2 IM survey.}
    \label{fig:omegade_lowz_highz}
\end{figure*}

\subsection{Intensity mapping: effect of foreground treatment}

As shown in Fig.~\ref{fig:obs_errs}, the performance of 21cm intensity mapping experiments depends to a large extent on how many Fourier modes are lost to foreground contamination. There is typically a difference of an order or magnitude or more in the constraints on $D_A$ between the most optimistic (no wedge) and pessimistic (horizon wedge) foreground removal scenarios for all of the IM experiments, and a little less than an order of magnitude for $H$. Fig.~\ref{fig:omegade_hirax} shows how this translates into constraints on $\rho_{\rm DE}(z)$ for the Tracker model, using HIRAX as an example. The difference in performance across the range of foreground removal assumptions is much less pronounced than for the more direct observables $D_{\rm A}$(z) and $H(z)$; the intermediate ($3\times$ primary beam wedge) case results in only mildly worse constraints than the optimistic no-wedge case for example.

The pessimistic horizon wedge case does perform substantially worse, offering little improvement over CMB + LSS on the constraints for models where $\rho_{\rm DE}(z)$ decays with increasing redshift (see the lower dashed red line in Fig.~\ref{fig:omegade_hirax}). This is because the HIRAX horizon wedge case provides only few-percent constraints on $H(z)$ and $D_{\rm A}(z)$, which is comparable to existing LSS surveys. These constraints are mostly at higher redshift than the existing LSS constraints however, where (as per Fig.~\ref{fig:model_wzhz}) the deviations from the fiducial $\Lambda$CDM model are smaller. Unless the HIRAX constraints are significantly tighter (as is the case for the no-wedge and $3\times$ PB wedge scenarios), there is therefore little additional constraining power to be gained from the HIRAX measurements, and the results are similar to the CMB + LSS-only case at $z \lesssim 2$. The higher-$z$ measurements do have more of an impact for models where the dark energy density is growing into the past however.

\subsection{Low redshift vs. high redshift constraints}
\label{sec:lowzhighz}

In this section we compare the relative performance of low- and high-redshift surveys in constraining $\rho_{\rm DE}(z)$. While the Tracker model has the flexibility to essentially decouple the values of the dark energy equation of state at low and high redshift, the observables all depend on integrals of this quantity, and so low redshift surveys do provide some information on what is happening in the high redshift regime and vice versa. The question is whether significant additional information on dark energy models can be gained from future high-redshift surveys (e.g. 21cm intensity mapping experiments or dedicated spectroscopic follow-up of LSST galaxies), or if low-$z$ surveys (most likely larger spectroscopic galaxy surveys) are likely to be sufficient.

The left panel of Fig.~\ref{fig:omegade_lowz_highz} shows how constraints from a cosmic variance-limited spectroscopic galaxy redshift survey covering around half the sky from $z = 0.1 - 1.5$ are affected by the addition of higher-redshift information. The CV-limited survey clearly offers a large improvement over existing CMB + LSS constraints, particularly at the very lowest redshifts. Combining it with HETDEX ($z = 1.9 - 3.5$) or a high-redshift version of HIRAX ($z = 2.0 - 6.1$) makes practically no difference to the constraints however; within the context of the Tracker model, all of the relevant parameters are already so well-constrained by the low-redshift survey that these higher-$z$ surveys are not sensitive enough to provide any further useful information. Note that the CV-limited survey does not offer a significant improvement over DESI (e.g. compare with Fig.~\ref{fig:omegade_galsurv}), and so it is realistic that similar constraints might be obtained in the near future when DESI data are released. Note that the combination of DESI with the CV-limited survey (e.g. a `DESI Southern Hemisphere') results in little additional improvement.

Combination with the Stage 2 survey (with $3\times$ PB foreground wedge removal) does make a significant difference however. Stage 2 offers around an order of magnitude improvement in $H$ and $D_A$ constraints over the same redshift range as the high-$z$ version of HIRAX, which translates to a factor of $\sim 3$ improvement when combined with the CV-limited survey, compared to that survey alone. The right-hand panel of Fig.~\ref{fig:omegade_lowz_highz} shows how this compares with Stage 2 only; while the combination of the CV-limited + Stage 2 surveys does offer the tightest constraints, Stage 2 already realises most of the improvement on its own. The main improvement to be gained from combining the two is around $z \approx 0.5$ for models with a decaying dark energy density into the past; the Stage 2-only case allows more freedom there due to its lack of low-redshift bins.

\begin{figure}
	\includegraphics[width=1.0\columnwidth]{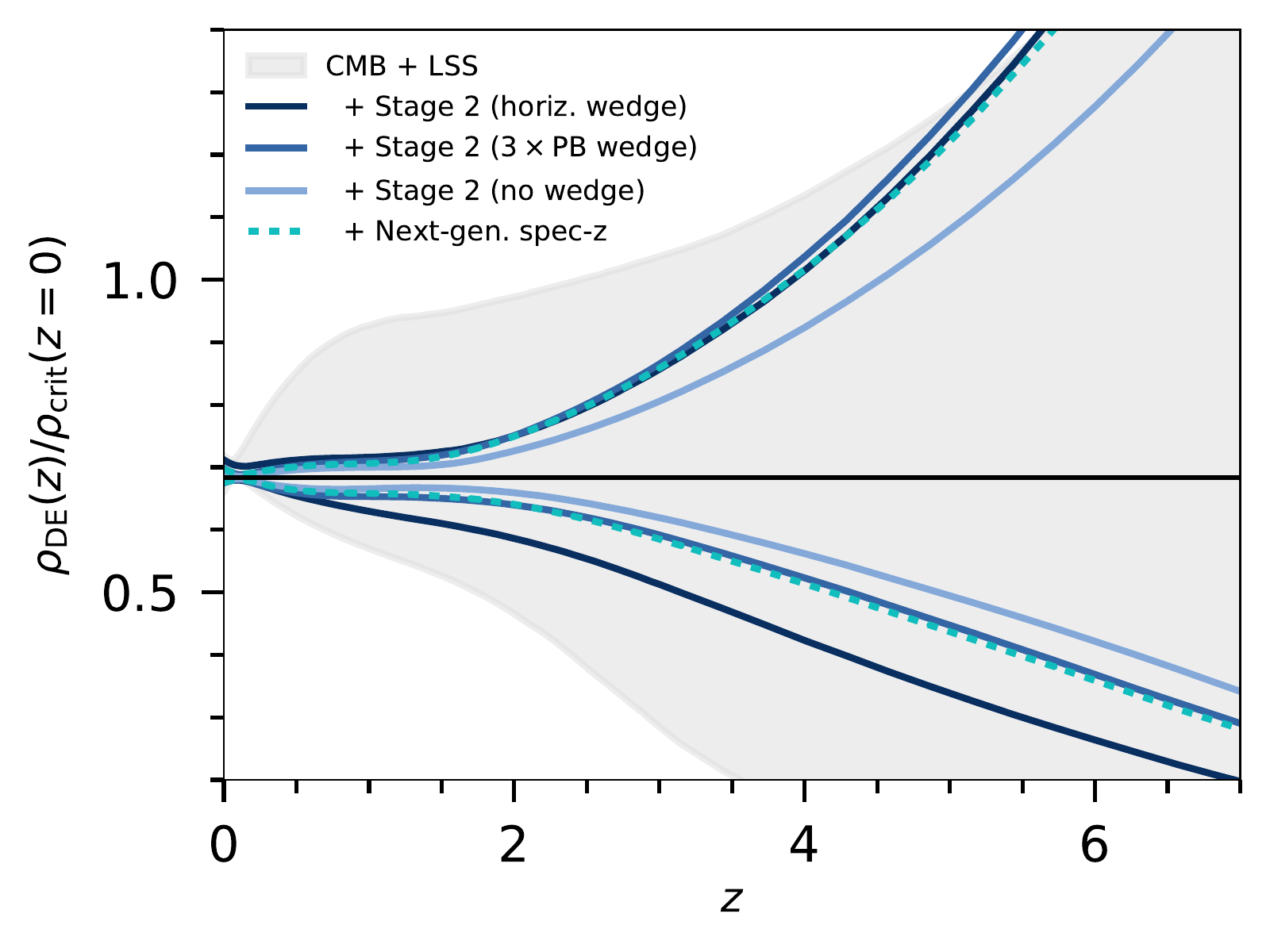}
    \caption{Forecast 95\%CL constraints on $\rho_{\rm DE}(z)$ for the Tracker model, comparing the next-generation spectroscopic galaxy redshift survey with the Stage 2 intensity mapping experiment under different foreground removal assumptions.}
    \label{fig:highzcomp}
\end{figure}

Finally, we compare the Stage 2 intensity mapping experiment with a high-redshift spectroscopic galaxy survey in Fig.~\ref{fig:highzcomp}. From Fig.~\ref{fig:obs_errs}, it is clear that the galaxy survey should have quite similar performance to the Stage 2 experiment with $3\times$ primary beam wedge foreground removal; the constraints on $H(z)$ are very similar over the broad overlapping redshift range, while the constraints on $D_A(z)$ are a factor of $\sim 2$ better due to the degradation of the angular resolution of the intensity mapping survey with increasing wavelength. Stage 2 does cover a broader redshift range however. This similarity is borne out by Fig.~\ref{fig:highzcomp}, which shows that Stage 2 ($3\times$ PB wedge) and the galaxy survey produce very similar constraints, with the galaxy survey performing only fractionally better at lower redshift. The intensity mapping survey does return significantly better constraints under the highly optimistic `no wedge' foreground removal assumptions however (although note that this is partially due to its larger assumed survey area).

Our conclusion from this analysis is therefore that higher-$z$ intensity mapping surveys and galaxy surveys (if sufficiently sensitive and free of systematics) are broadly comparable, and should be capable of putting generally more stringent constraints on a broad class of dark energy models than an equivalent lower-$z$ spectroscopic galaxy survey -- although at the expense of being able to constrain some low-redshift phenomenology as well.

\subsection{Comparison of Mocker and Tracker models}

In addition to the phenomenological Tracker model, we also consider the more constrained Mocker model, as discussed in Sect.~\ref{sec:demodels}. This also exhibits a transition in $w(z)$, which begins immediately at $z=0$ (rather than having a tunable location; c.f. the $z_c$ parameter in the Tracker model). Its equation of state is also constrained to be $w \ge -1$, which means that $\rho_{\rm DE}(z)$ can only increase into the past (or stay constant in the limiting case, $w=-1$).

\begin{figure}
	\includegraphics[width=1.0\columnwidth]{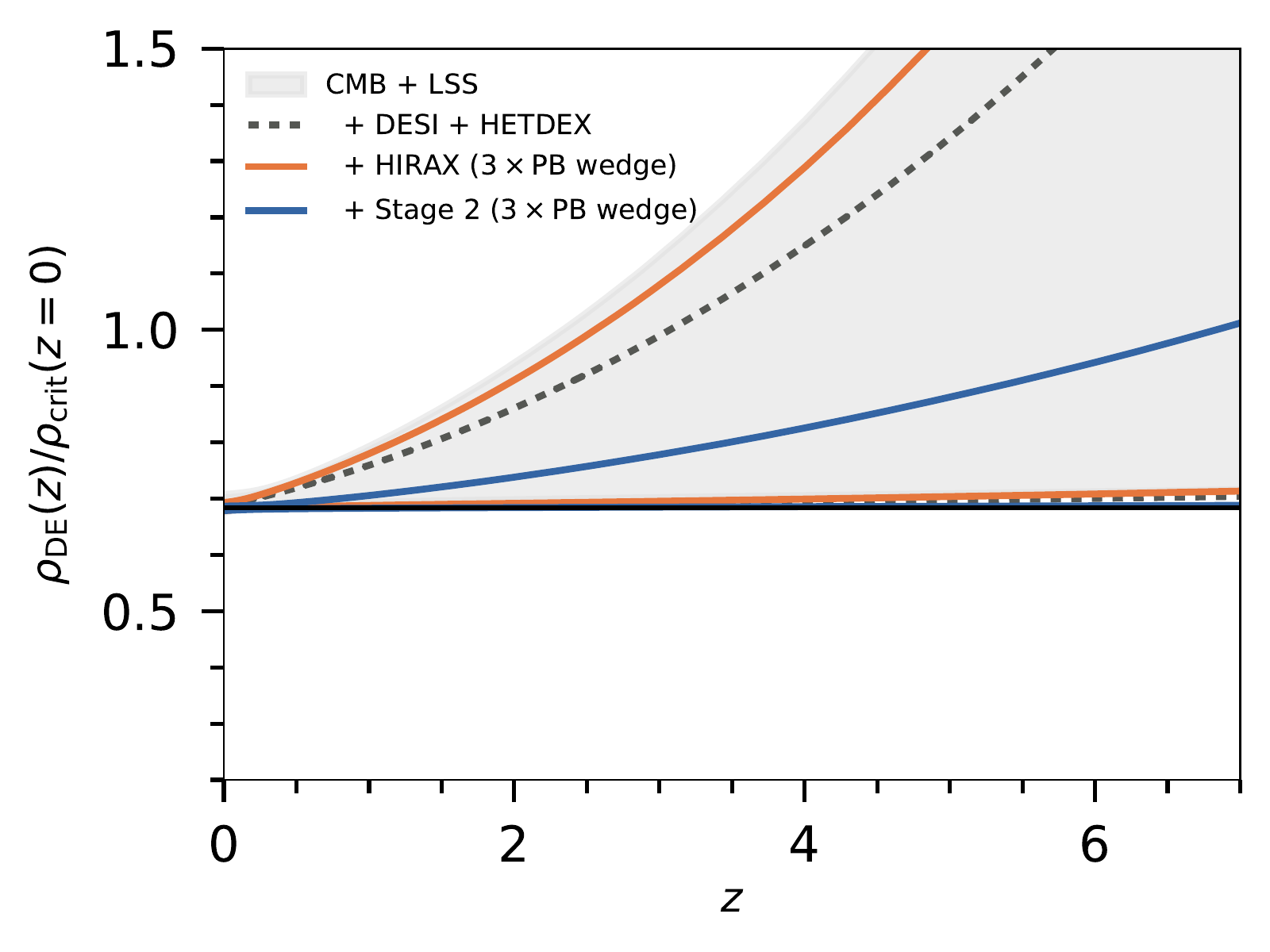}
    \caption{Forecast 95\%CL constraints on $\rho_{\rm DE}(z)$ for the Mocker model, for the same set of surveys as Fig.~\ref{fig:omegade_galsurv}. The hard bound on $w \ge -1$ in this model requires that $\rho_{\rm DE}(z)$ grows into the past.}
    \label{fig:mockertracker}
\end{figure}

Fig.~\ref{fig:mockertracker} shows constraints on the Mocker model for the same set of surveys that were shown in Fig.~\ref{fig:omegade_galsurv}. In this case, there is a more significant difference in the behaviour of the DESI + HETDEX constraints compared with HIRAX. This is due to the more constrained Mocker parametrisation; both low- and high-redshift constraints are capable of strongly constraining the $C$ parameter, which governs the behaviour of $w(z)$ across the entire redshift range. As illustrated in Fig.~\ref{fig:model_wzhz}, the expansion rate is slightly more sensitive to the value of $C$ at lower redshift, and the lower-$z$ constraints from DESI are stronger than the intermediate-redshift constraints from HIRAX (albeit over a narrower redshift range), hence the better performance of DESI + HETDEX in this case. The Stage 2 constraints are considerably better than DESI + HETDEX (more so than for the Tracker model), again because of the global dependence of $w(z)$ on the $C$ parameter.

\begin{figure}
	\includegraphics[width=1.0\columnwidth]{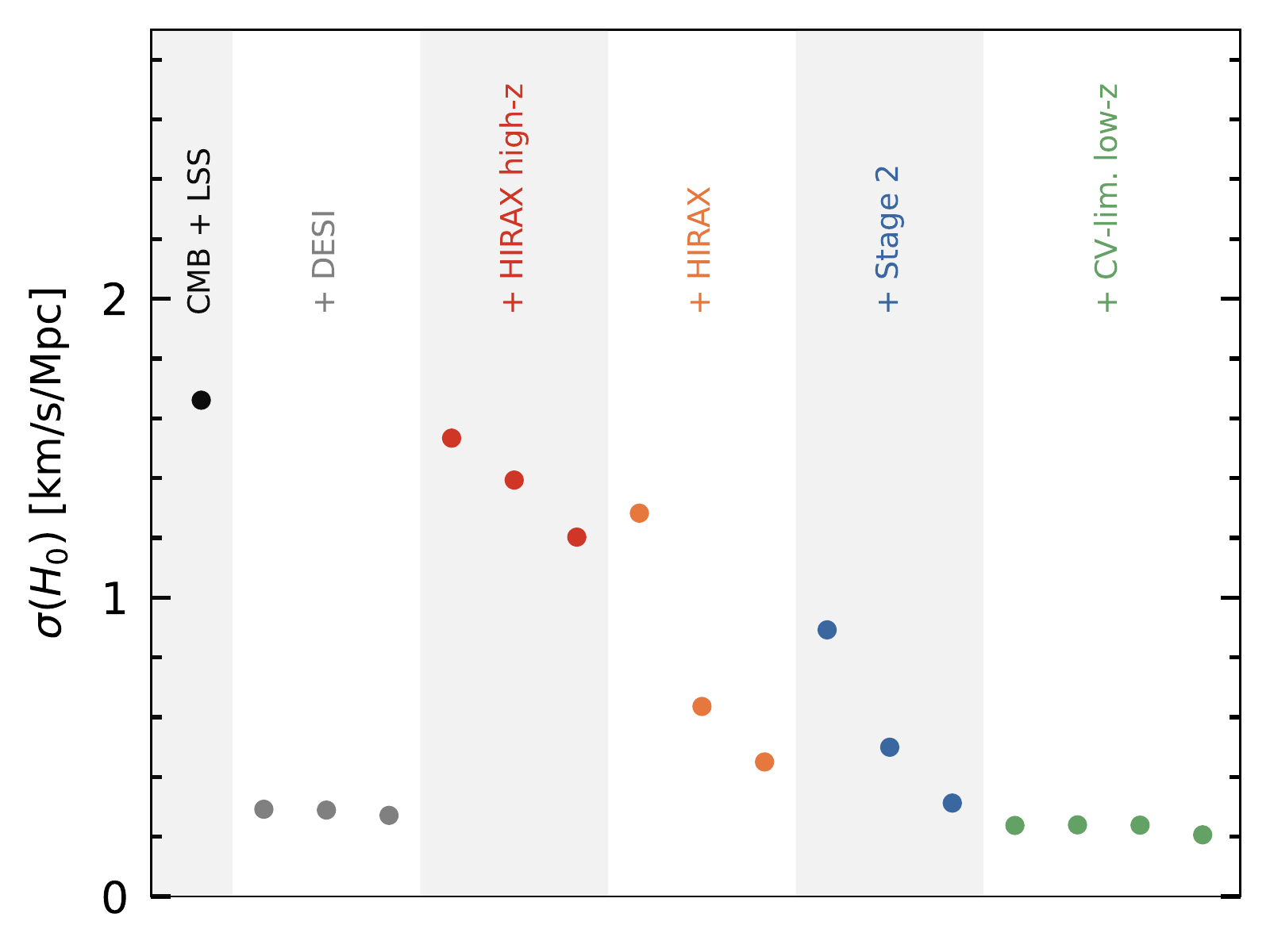}
    \caption{Constraints on the Hubble parameter $H_0$ (68\% CL) assuming the Tracker model, for a range of experiments. From left to right, the experiments are: CMB + LSS only; including DESI (on its own, with HETDEX, and with HIRAX $3\times$PB wedge); including HIRAX high-z (horizon, $3\times$PB, and no wedge); including HIRAX (same three wedge types); including Stage 2 (same three wedge types); and including CV-limited low-z (on its own, with HETDEX, with HIRAX high-z $3\times$PB wedge, and with Stage 2 $3\times$PB wedge).}
    \label{fig:tanh_hubble}
\end{figure}

In terms of the confidence intervals on $\rho_{\rm DE}(z)$, it is notable that the Mocker model exhibits only mildly tighter constraints at low redshift, despite having significantly less flexibility than the Tracker model. This can be seen by comparing $\rho_{\rm DE}(z)$ at fixed redshift between Figs.~\ref{fig:omegade_galsurv} and \ref{fig:mockertracker}. For example, for the HIRAX curve at $z=2$, the width of the 95\% CL contour is $\Delta \rho_{\rm DE}/\rho_{{\rm crit},0} \approx 0.2$ for Mocker and $\approx 0.25$ for the Tracker model. The difference is more noticeable at higher-$z$; for example, for Stage 2 the corresponding values at $z=4$ are $\approx 0.1$ (for Mocker) and $0.5$ (for Tracker).

\subsection{Model parameter constraints}
\label{sec:cosmoparams}

In the above, we have focused on constraints on $\rho_{\rm DE}(z)$. While these are technically model-dependent, the phenomenological $\tanh$ model for $w(z)$ is reasonably general, and contains several other commonly-used parametrisations (like the CPL $w_0-w_a$ parametrisation) as limits. The results above should therefore give a broadly conservative picture of how well a variety of dark energy models can be constrained.

A key parameter that depends on the dark energy model is the Hubble parameter, $H_0$. Discrepancies between measurements of $H_0$ using different methods have led to suggestions that the $\Lambda$CDM model may need to be extended in order to provide an explanation for this anomaly. Dark energy models have been studied as a potential solution to this problem \citep{Karwal:2016vyq, Poulin:2018cxd, Keeley:2019esp, Lin:2019qug, Knox:2019rjx}, amongst many others. While we do not study the Hubble discrepancy specifically in this paper, we can comment on the constraints on $H_0$ that can be achieved by various combinations of surveys after marginalising over a Tracker-type dark energy model.

\begin{figure*}
	\includegraphics[width=2\columnwidth]{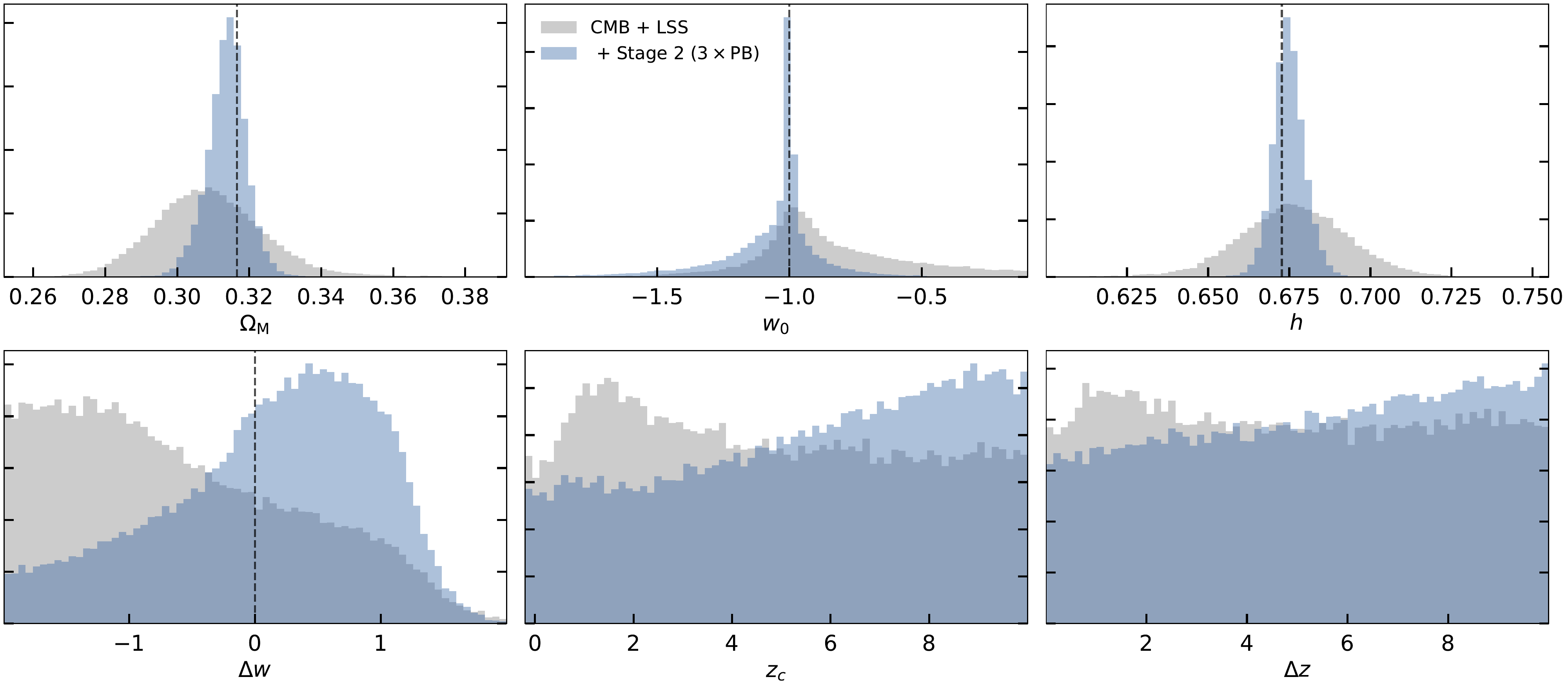}
    \caption{Marginal distributions for 6 of the 7 parameters in the Tracker model, for two combinations of survey data. Dashed vertical lines show fiducial parameter valiues where appropriate.}
    \label{fig:marginals}
\end{figure*}

Fig.~\ref{fig:tanh_hubble} shows the marginal uncertainty on $H_0$ for several combinations of experiments, now at the 68\% CL (rather than 95\% as in previous figures). All of them measure $H_0$ in a similar way to the existing large-scale structure surveys that appear to agree with the `high-redshift' determination of $H_0$ from Planck (more so than the `low-redshift' determination from Type Ia supernovae, distance ladder measurements, and strong gravitational lenses). If the $H_0$ discrepancy persists, these surveys would therefore be contributing to reducing the uncertainty on this one particular type of $H_0$ measurement, without necessarily saying anything about the low-redshift measurements. The Tracker parametrisation used here would allow scenarios with rapid low-redshift transitions in the equation of state of dark energy to be tested though, and would also contribute to constraints on early dark energy models \citep[although see the recent analysis by][]{Hill:2020osr}.

As shown in Fig.~\ref{fig:tanh_hubble}, a higher-$z$ HIRAX configuration would do comparatively little to improve the $H_0$ constraints. The standard HIRAX configuration does better, improving the uncertainty from $\sigma(H_0) \approx 1.6$ km/s/Mpc from existing CMB + LSS constraints to $\approx 0.6$ km/s/Mpc in the $3\times$ PB wedge foreground removal case. This is improved to $0.5$ km/s/Mpc for Stage 2 with the same foreground removal assumptions. In comparison, DESI should already be able to achieve $\sigma(H_0) \approx 0.3$ km/s/Mpc, with the CV-limited low-$z$ achieving a little over $0.2$ km/s/Mpc. A slight improvement can be gained by combining the CV-limited low-$z$ experiment with Stage 2 (or alternatively using Stage 2 on its own in a no-wedge scenario), but the change from the DESI result is quite small. Under the assumptions of our analysis, then, DESI is already likely to offer the most decisive $H_0$ measurement using this particular method (although further improvements are possible with the addition of future CMB experiments and weak gravitational lensing surveys).

Fig.~\ref{fig:marginals} shows the 1D marginal constraints on the Tracker model parameters for the CMB + LSS-only data, compared with CMB + LSS + Stage 2 (assuming $3\times$ PB foreground wedge removal). Constraints on standard cosmological parameters such as $\Omega_{\rm M}$ and $h$ are much improved by adding Stage 2, which is a sign that the dark energy parametrisation we have adopted here is not overly flexible. Most of the dark energy parameters are poorly constrained in both cases however. The exception is $w_0$, which is constrained by Stage 2 to be in the range $[-1.60, -0.72]$ (95\% CL) in this example. 

As discussed in Sect.~\ref{sec:mcmc}, our fiducial model has $w=-1$, which maps to a subspace of Tracker models where $\Delta w = 0$, in which case $z_c$ and $\Delta z$ should be completely unconstrained. This is indeed what we see in Fig.~\ref{fig:marginals}, although the shape of the marginal distributions for these parameters does change when Stage 2 is added, presumably as previously viable models that are more distant from this subspace are ruled out. Still, these parameters fill their prior bounds in both cases, confirming that they are unconstrained.

Another point of interest in Fig.~\ref{fig:marginals} is the marginal distribution for $\Delta w$. Recall that $\Delta w = w_\infty - w_0$. For the CMB + LSS constraints, based entirely on current data, a skew towards negative values of $\Delta w$ is observed, although this is not statistically significant. This can be compared with the mild preference for values of $w_a < 0$ in Fig.~30 of \cite{2018arXiv180706209P} (e.g. for the combination of Planck, LSS, and supernova data). In Tracker models with a broad transition (large $\Delta z$) and moderately small value of $z_c$, the equation of state approximates a linear function at low to intermediate redshifts; writing it in the form $w(z) = w_0 + w_a(1 - a)$, the transition height $\Delta w \approx w_a$, and so we see that the mild preference for $\Delta w < 0$ is likely another manifestation of the slight preference already seen by Planck.

\begin{figure*}
	\includegraphics[width=1.0\columnwidth]{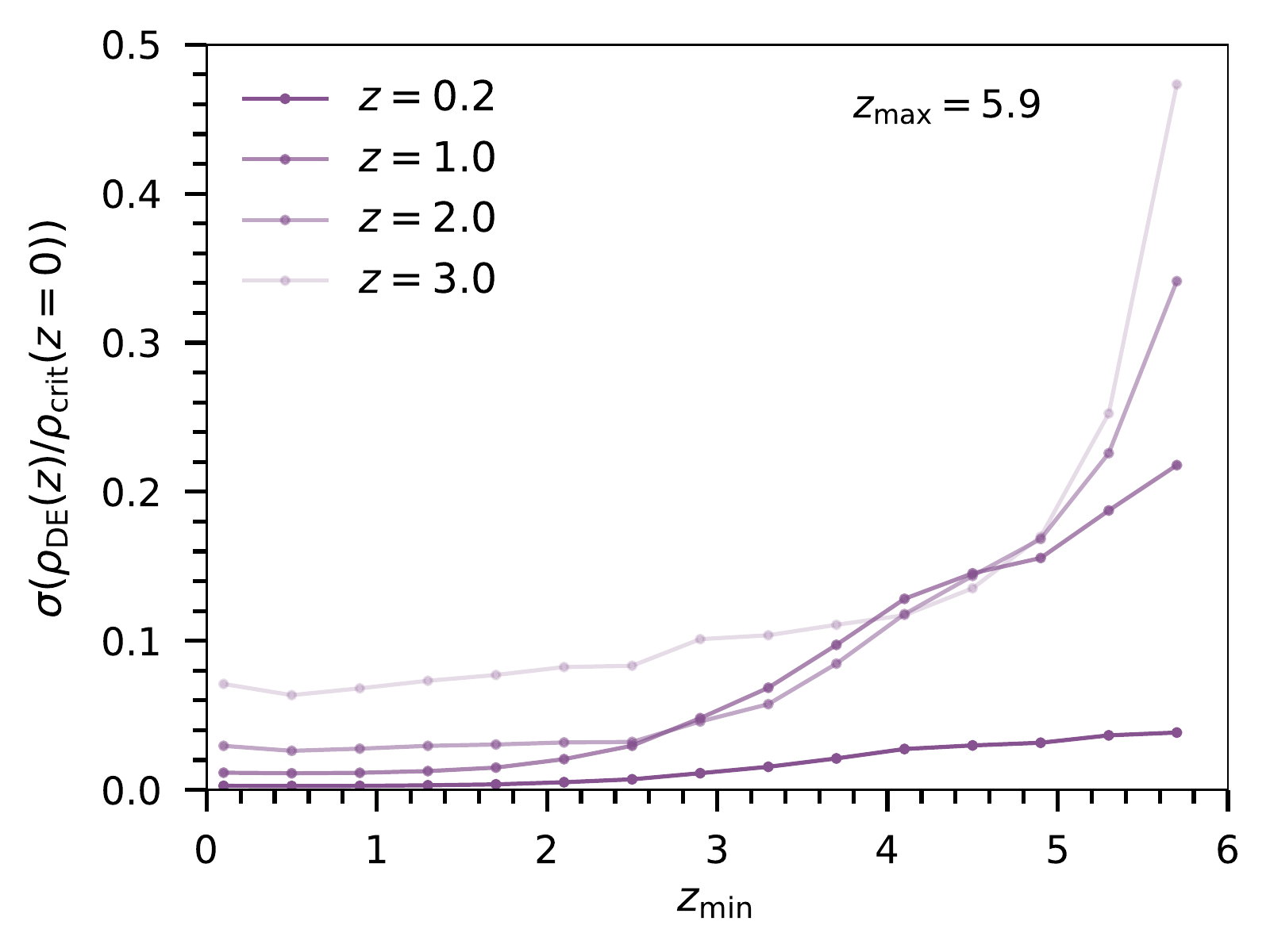}
	\includegraphics[width=1.0\columnwidth]{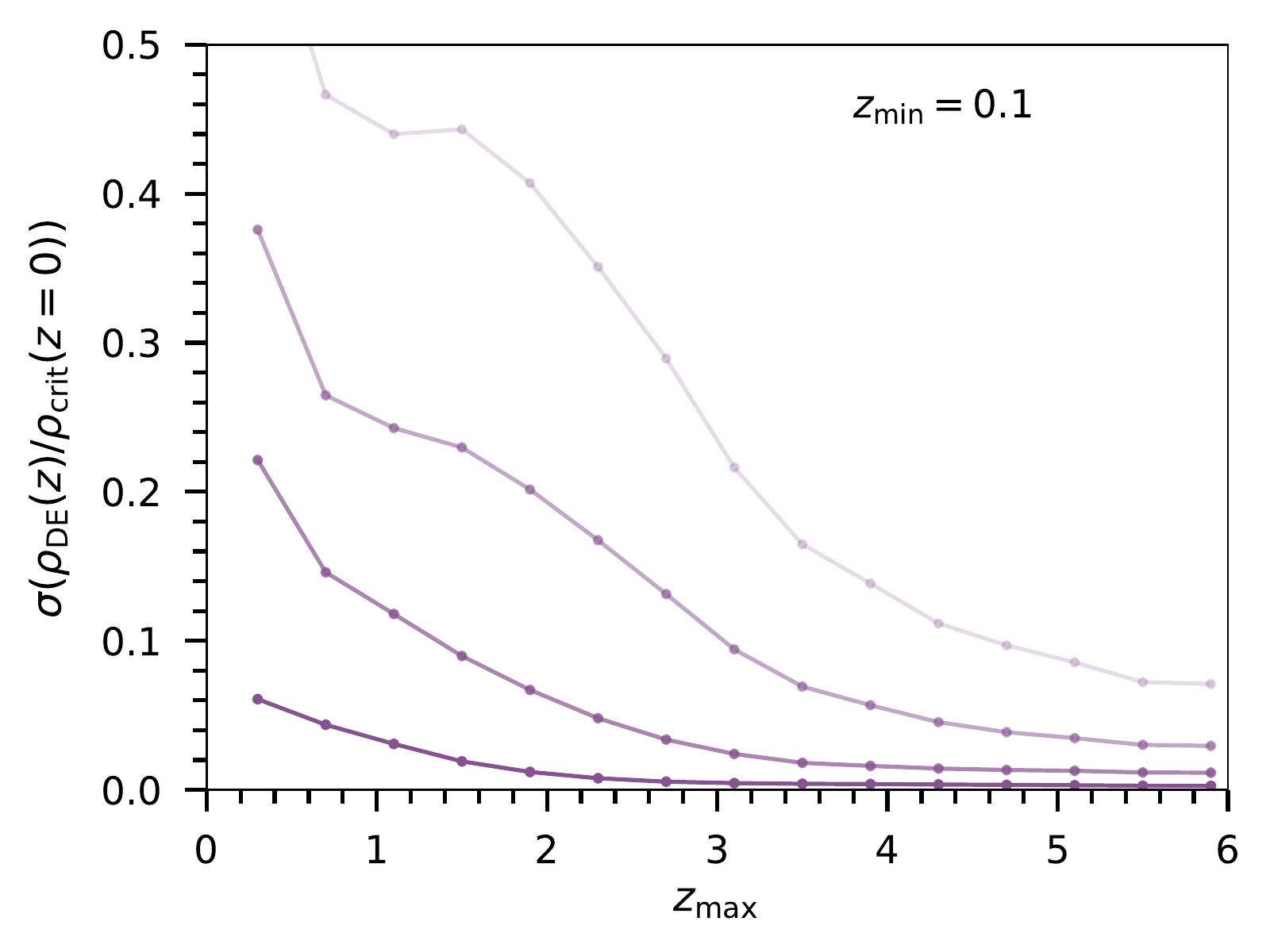}
    \caption{Forecast constraints on $\rho_{\rm DE}(z)$ in the Tracker model for a sample variance-limited survey (+ CMB + LSS) with varying upper and lower redshift limits for the survey. Each curve denotes the width of the 95\% CL interval of $\rho_{\rm DE}(z) / \rho_{\rm crit}(z=0)$ at a given redshift. ({\it Left panel:}) Forecast constraints for a fixed $z_{\rm min} = 0.1$ and a varying $z_{\rm max}$. ({\it Right panel:}) Forecast constraints for a fixed $z_{\rm max} = 5.9$ and a varying $z_{\rm min}$. Note that the number of bins is increased in steps of 2 in these plots, rather than one bin at a time.}
    \label{fig:omegade_cvallz}
\end{figure*}

\subsection{Optimal redshift window for future dark energy studies}
\label{sec:zlimits}

In this section we study whether there is an optimal survey redshift range for constraining the Tracker parametrisation. We focus our attention on the choice of the minimum and maximum redshift extent of the survey, $z_{\rm min}$ and $z_{\rm max}$, assuming a galaxy redshift survey with fixed survey area of 14,000 deg$^2$ (comparable to other galaxy surveys) and very high number density so that the survey is sample variance-limited in each redshift bin. We also fix the width of the redshift bins to $\Delta z = 0.2$, and include the existing CMB + LSS data in the likelihood for our predictions. Finally, we assume that the galaxy bias evolves with redshift as $b(z) = \sqrt{1+z}$. Our target observable is again the dark energy density as a function of redshift, $\rho_{\rm DE}(z)$.

Fig.~\ref{fig:omegade_cvallz} shows the resulting forecasts for a survey with a fixed maximum redshift but variable minimum redshift (left panel), and vice versa (right panel). These correspond to scenarios where a nominally high-redshift survey is extended to progressively lower redshifts (left panel), and a low-redshift survey is extended to progressively higher redshift (right panel).

For redshifts $z \ge 1$, gains are made more rapidly by starting at high redshift and decreasing $z_{\rm min}$. The point of diminishing returns is reached around $z_{\rm min} \sim 3$, with only small additional improvements gained by extending the redshift range lower. Interestingly, the constraints in all three redshift bins at $z \ge 1$ improve at a similar rate with decreasing $z_{\rm min}$ until this point, with the curves becoming flatter and more differentiated for $z_{\rm min} \lesssim 3$.

In contrast, starting at low redshift and increasing $z_{\rm max}$, the curves for the different redshift bins remain well-differentiated, with a much gentler improvement that only reaches a point of diminishing returns at $z \gtrsim 4 - 4.5$.

Taken in isolation, these results make a case for preferring new high redshift surveys over low redshift ones. Considering the four redshifts shown in Fig.~\ref{fig:omegade_cvallz}, one can learn about as much about the evolution of $\rho_{\rm DE}$ by surveying the range $z = [3.2, 6.0]$ as $z = [0.1, 4.6]$. This ignores important factors that weigh in both directions however, such as the relative technical difficulty of targeting higher redshifts, and the number of existing/imminent datapoints at lower redshift. It is also a model-dependent conclusion that depends on our choice of parametrisation and fiducial model ($w = -1 = {\rm const}$). Nevertheless, these results show that there is a case to be made for high-$z$ observational studies of dark energy.

\section{Conclusions}
\label{sec:conclusions}

Current observations are so far consistent with a Cosmological Constant (CC) being the dominant driver of accelerated expansion at late times. If the acceleration is instead caused by a dark energy fluid, we have few clues about whether and how this would differ observationally from a CC -- many dark energy models contain the CC as a limit, and it is hard to make convincing arguments (e.g. on grounds of naturalness or consistency) for why we should expect new fields to follow anything other than slowly-rolling, potential-dominated trajectories. Indeed, in many models that have been studied, CC-like behaviour is found to be an attractor. Without a universally agreed-upon measure over the space of possible dark energy theories to tell us which models are more likely to be realised in nature, it seems unlikely that we can make headway beyond simply measuring the cosmic background expansion ever more precisely and thus progressively ruling out alternatives.

That said, it is possible to {\it assume} a measure and then search for `generic' behaviours in broad classes of dark energy models. While selecting a measure by fiat hardly guarantees universality, it can at least provide instructive `what-if' scenarios. In this paper, we have used the results of such an approach to argue that there may be interesting dark energy phenomenology at higher redshifts than are normally probed by late-time experiments, back in the matter-dominated era. Our argument is based on the observation that generalised single scalar field theories -- the Horndeski class of models -- admit couplings to the matter sector that cause tracking-type behaviour of the dark energy equation of state. This tends to lead to a transitioning behaviour in $w(z)$ beginning in the matter-dominated epoch, $z \gtrsim 2$, which we have attempted to capture using a phenomenological Tracker model and a more particular `Mocker' toy model. The detection of such a transition feature in the equation of state would be a compelling signature of physics beyond the CC, and so is an important phenomenon to search for even if low-redshift constraints remain consistent with $w = -1$.

We note that the Fisher forecasting methodology we used to make predictions for future experiments (see Sect.~\ref{sec:fisher}) is optimistic in a number of respects. First, we have used the full broadband shape of the power spectrum to derive constraints, rather than the more conservative BAO-only approach that is currently taken by most galaxy surveys. Recent advances in modelling the power spectrum \citep[including nonlinear effects, baryonic effects etc.; see e.g.][]{2016arXiv160200674B, 2016JCAP...05..027F, Sprenger:2018tdb, Schneider:2018pfw, Ivanov:2019pdj, Nishimichi:2020tvu} suggest that it is not overly optimistic to assume that the broadband power spectrum can be modelled accurately out to the mildly non-linear scale cut we have applied in this paper, at least for galaxy surveys such as DESI.

For the intensity mapping experiments, there are additional challenges in the form of bandpass calibration uncertainties and other calibration artifacts that could conceivably distort the broadband power spectrum. We have not modelled these, and the level at which these can be controlled in large IM surveys is not yet understood. We have also neglected uncertainties on the HI physics, such as the mean brightness temperature as a function of redshift, $T_b(z)$, which is poorly known at present \citep{2017MNRAS.471.1788C, 2018ApJ...866..135V}. If its functional form is unconstrained a priori, there is a hard degeneracy between $T_b(z)$ and parameters such as $f\sigma_8$, although this can be ameliorated by adding additional information, e.g. from nonlinear scales \citep{2018JCAP...05..004O, 2019arXiv190207147C}. The single most important systematic for IM surveys is foreground contamination however \citep{Seo:2015aza}, and the problems that the large dynamic range between foregrounds and cosmological signal cause for calibration. We have presented a range of foreground contamination scenarios, from very optimistic (perfect foreground cleaning) to pessimistic (full removal of the foreground wedge), and so, modulo any as-yet undiscovered `show-stopper' systematics, we are confident that our analysis here brackets the performance that will be achieved by these experiments in reality.

We have shown how existing constraints allow a broad range of behaviours of the redshift evolution of dark energy. Future experiments at higher redshifts can improve upon current constraints on the redshift evolution of $\rho_{\rm DE}$ by a factor of a few, with low- to intermediate-redshift experiments like a DESI spectroscopic galaxy survey and a HIRAX 21cm intensity mapping survey able to realise about half of the expected gain in precision over the next few years. These two surveys reach similar levels of precision through very different means -- DESI achieves a larger effective volume (and therefore significantly smaller errorbars) at $z \lesssim 1.5$, while HIRAX compensates for losing a substantial fraction of modes to foreground filtering by extending much deeper, out to $z \simeq 2.4$.

Following DESI, we also considered a cosmic variance-limited galaxy survey over half the sky out to $z_{\rm max} = 1.5$, but found little improvement in constraints. Instead, substantial improvements on the constraints on dark energy evolution -- even at lower redshifts -- will likely require a wide and deep survey from $z \gtrsim 2$ up to as high a redshift as $z \sim 6$. This type of survey was modelled by the high-$z$ HIRAX and Stage 2 intensity mapping configurations in this paper, as well as a next-generation spectroscopic galaxy survey that would follow-up the LSST galaxy sample at $z \gtrsim 2$, along similar lines to proposals such as FOBOS \citep{2019BAAS...51g.198B}, Maunakea Spectroscopic Explorer \citep{2019BAAS...51g.126M}, MegaMapper \citep{Schlegel:2019eqc}, and SpecTel \citep{2019BAAS...51g..45E}. The sensitivity of a HIRAX-like instrument will not be enough to result in significant gains, and so for intensity mapping an instrument of a similar scale as the Stage 2 proposal of \cite{Ansari:2018ury} (tens of thousands of receiving elements) will be required. Recent proposals for such an instrument include the Packed Ultra-wideband Mapping Array \citep[PUMA;][]{Bandura:2019uvb}, which could begin operations in the early 2030s and survey $0.3 \lesssim z \lesssim 6.1$, and the Canadian Hydrogen Observatory and Radio transient Detector \citep[CHORD;][]{Liu:2019jiy}, a smaller array that extends only to $z \sim 3.7$. We also found that a high-$z$ galaxy survey following the `idealised' specification in \cite{Ferraro:2019uce} would offer similar (fractionally better) performance than Stage 2 under the intermediate `$3\times$ primary beam wedge' foreground removal treatment.

In Sect.~\ref{sec:cosmoparams} we examined how the various combinations of experiments would affect cosmological parameter constraints. We found that, under our assumptions, DESI should already provide close to the best possible constraints on $H_0$ from galaxy clustering, with very little improvement possible from subsequent experiments. We also studied the constraints that can be achieved on the parameters of the flexible Tracker parametrisation of the dark energy equation of state, finding that significant freedom remains in parameters such as $w_0$ even for the most powerful combinations of experiments. This is a model-dependent statement however, and we have shown that significantly improved constraints on the time evolution of the dark energy density, $\rho_{\rm DE}(z)$, can be achieved by the same set of experiments. We also briefly note that we did not consider the impact of these experiments on neutrino mass constraints; these were studied in a similar context by \cite{2018JCAP...05..004O}, who found that significant improvements on $\sum m_\nu$ are possible if large-scale structure surveys are extended to higher redshift.

Finally, in Sect.~\ref{sec:zlimits} we considered an idealised sample variance-limited galaxy survey, changing its upper and lower redshift limits to try and identify an `optimal' redshift range for constraining the redshift evolution of dark energy, again under the assumption of our Tracker parametrisation for $w(z)$. We found that a survey over $z = [3.2, 6.0]$ should constrain the evolution of $\rho_{\rm DE}(z)$ about as well as a survey over $z = [0.1, 4.6]$, suggesting that there is now more information to be gained by building high redshift galaxy/intensity mapping surveys than low redshift ones. This conclusion is obviously sensitive to the detailed specifications of the surveys in question, and particularly the feasibility of going out to high redshift (e.g. in terms of detectable source populations and various systematics), but nevertheless suggests that spectroscopic surveys at $z \gtrsim 2$ are a promising future direction for the observational study of dark energy.

\section*{Acknowledgements}

We are grateful to Y.~Akrami, T.~Baker, E.~Castorina, and the Cosmic Visions 21cm working group for useful discussions. PB acknowledges support from RAL at UC Berkeley. We acknowledge use of the following software: {\tt emcee} \citep{2013PASP..125..306F}, {\tt matplotlib} \citep{matplotlib}, {\tt numpy} \citep{numpy}, and {\tt scipy} \citep{2020SciPy-NMeth}.

\balance

\bibliographystyle{mnras}
\interlinepenalty=10000
\bibliography{earlyde}

\end{document}